\documentclass[apj]{emulateapj}
\usepackage{epstopdf}
\usepackage{color}
\usepackage{hyperref}
\usepackage{ulem}

\shorttitle{Infall time and Mass loss in Phase-Space}
\shortauthors{Rhee et al.}

\begin{document}

%%%%%%%%%%%%%%%%%%%%%%%%%%%%%%%%%%%%%%%%%%%%%%%%%%%%%%%%%%%%%%%%%%%%%%%%%%%%%%%%%%%%%%%%%%%%%%%%%%%%%%%%%%%%%%%%%%%
%%%%%%%%%%%%%%%%%%%%%%%%%%%%%%%%%%%%%%%%	Title	%%%%%%%%%%%%%%%%%%%%%%%%%%%%%%%%%%%%%%%%%%%%%%%%%%%%%%%%%%%%%%%%%%%%%%
%%%%%%%%%%%%%%%%%%%%%%%%%%%%%%%%%%%%%%%%%%%%%%%%%%%%%%%%%%%%%%%%%%%%%%%%%%%%%%%%%%%%%%%%%%%%%%%%%%%%%%%%%%%%%%%%%%%
\title{Phase-space Analysis in the Group and Cluster environment: Time since Infall and Tidal Mass Loss}

\author{Jinsu Rhee$^{1}$, Rory Smith$^{1}$, Hoseung Choi$^{1}$, Sukyoung K. Yi$^{1}$, Yara Jaff\'{e}$^{2}$, Graeme Candlish$^{3}$ and Ruben S\'{a}nchez-J\'{a}nssen$^{4}$}
\affil{$^1$Department of Astronomy and Institute of Earth-Atmosphere-Astronomy, Yonsei University, Seoul 03722, Korea; jinsu.rhee@yonsei.ac.kr\\
$^2$European Southern Observatory, Alonso de Cordova 3107, Vitacura, Casilla 19001, Santiago de Chile, Chile\\
$^3$Universidad de Valparaíso, Blanco 951, Valparaíso, Chile\\
$^4$UK Astronomy Technology Centre,  Royal Observatory,
Blackford Hill, Edinburgh EH9 3HJ, UK\\}

%%%%%%%%%%%%%%%%%%%%%%%%%%%%%%%%%%%%%%%%%%%%%%%%%%%%%%%%%%%%%%%%%%%%%%%%%%%%%%%%%%%%%%%%%%%%%%%%%%%%%%%%%%%%%%%%%%%
%%%%%%%%%%%%%%%%%%%%%%%%%%%%%%%%%%%%%%%%	Abstract	%%%%%%%%%%%%%%%%%%%%%%%%%%%%%%%%%%%%%%%%%%%%%%%%%%%%%%%%%%%%%%%%%%%
%%%%%%%%%%%%%%%%%%%%%%%%%%%%%%%%%%%%%%%%%%%%%%%%%%%%%%%%%%%%%%%%%%%%%%%%%%%%%%%%%%%%%%%%%%%%%%%%%%%%%%%%%%%%%%%%%%%

\begin{abstract}
Using the latest cosmological hydrodynamic N-body simulations of groups and clusters, we study how location in phase-space coordinates at $z$$=$$0$ can provide information on environmental effects acting in clusters. We confirm the results of previous authors showing that galaxies tend to follow a typical path in phase-space as they settle into the cluster potential. As such, different regions of phase-space can be associated with different times since first infalling into the cluster. However, in addition, we see a clear trend between total mass loss due to cluster tides, and time since infall. Thus we find location in phase-space provides information on both infall time, and tidal mass loss. We find the predictive power of phase-space diagrams remains even when projected quantities are used (i.e. line-of-sight velocities, and projected distances from the cluster). We provide figures that can be directly compared with observed samples of cluster galaxies and we also provide the data used to make them as supplementary data, in order to encourage the use of phase-space diagrams as a tool to understand cluster environmental effects. We find that our results depend very weakly on galaxy mass or host mass, so the predictions in our phase-space diagrams can be applied to groups or clusters alike, or to galaxy populations from dwarfs up to giants. 
\end{abstract}

\keywords{}

%%%%%%%%%%%%%%%%%%%%%%%%%%%%%%%%%%%%%%%%%%%%%%%%%%%%%%%%%%%%%%%%%%%%%%%%%%%%%%%%%%%%%%%%%%%%%%%%%%%%%%%%%%%%%%%%%%%
%%%%%%%%%%%%%%%%%%%%%%%%%%%%%%%%%%%%%%%%	Introduction	%%%%%%%%%%%%%%%%%%%%%%%%%%%%%%%%%%%%%%%%%%%%%%%%%%%%%%%%%%%%%%%%
%%%%%%%%%%%%%%%%%%%%%%%%%%%%%%%%%%%%%%%%%%%%%%%%%%%%%%%%%%%%%%%%%%%%%%%%%%%%%%%%%%%%%%%%%%%%%%%%%%%%%%%%%%%%%%%%%%%

\section[]{Introduction}
\label{sec:introduction}

The morphology-density relation \citep{Dr80} demonstrates that different types of galaxies prefer different environments. For example, late-type spiral galaxies tend to avoid the higher density environments, while early type galaxies tend to prefer them. This discovery prompted substantial investigations into the root cause of the relation. One possibility is that environmental mechanisms may be preferentially acting in higher density environments, halting star formation and/or transforming galaxies morphologically. 

Galaxies inside the dense environment of galactic clusters may be subjected to several types of environmental mechanisms simultaneously. The large mass of a cluster ($\sim$$10^{14-15}$~M$_\odot$) generates a deep potential well. Within this potential well, a multimillion Kelvin hot, ionized gas is contained, known as the intracluster medium. The deep gravitational potential drives high galaxy orbital velocities (typically $\sim$1000~km/s). As a result, cluster galaxies pass through the intracluster medium at high velocities, and the collision of the intracluster medium with the neutral, atomic gas disk of late-type cluster galaxies generates a ram pressure \citep{GG72}. This process, known as `ram pressure stripping', can strip away the atomic gas disk \citep{Abetal99,Chetal07,Jaetal07}, and may be responsible for converting some star forming late-type disk galaxies into red and dead, early-type galaxies \citep{Boetal08}. When ram pressure stripping is not sufficient to strip the atomic gas disk, it may still be strong enough to remove warm, ionized gas within the halo of a galaxy \citep{Be09,Mcetal08,Foetal08}. This process, known as `starvation', can cut off the inflow of fresh disk gas that cools out from the hot halo, causing the galaxy to slowly use up its remaining disk gas \citep{Laetal80}.

The deep potential well of the cluster can also generate strong tidal accelerations on cluster galaxies, which may truncate their dark matter haloes \citep{Lietal09,Gaetal04,Waetal08}, drive gas to the center, inducing a central starburst \citep{BV01}, trigger bar instabilities, and enhanced spiral structure \citep{Loetal16,Seetal17}. Furthermore, there maybe $\sim$1000 other members within the cluster, each moving at high velocity. As a result, a cluster galaxy can be subjected to repeated high speed tidal encounters, combined with the effects of the overall cluster potential -- a process known as `harassment' \citep{Moetal98}. This process has been shown to be effective especially acting on extended, low surface brightness disks \citep{Moetal99,Gnetal03}.  The strength of mass loss from harassment has been shown to be highly sensitive to the type of orbit a galaxy follows in the cluster \citep{Maetal05,Smetal10,Smetal13,Smetal15}.

To realistically interpret populations of galaxies residing in clusters today, it may be necessary to consider environmental effects that act outside of the cluster. For example, some galaxies may have resided in a group of galaxies, and then fallen into the cluster within that group. Any environmental mechanisms that act in groups could have previously influenced that galaxy before entering the cluster, a process known an `pre-processing' \citep{Mi2004}. A non-negligible fraction of cluster galaxies have spent some time in a group environment, prior to cluster infall \citep{Mcetal09,Deetal12,Weetal13,Hoetal14}. Dedicated numerical simulations of the galaxy group environment demonstrate that tidal mass loss can be severe for group members \citep{Vietal12,Vietal14}. Identifying galaxies that have been pre-processed may not be easy, as they may have been removed from their groups by the cluster tides. 
Deep images were lately used to reveal merger features in cluster galaxies which can be attributed to merger events before they fell into the current cluster \citep{Shetal12,Yietal13}. Furthermore, the combination of the group environment with the cluster environment may temporarily drive even stronger environmental effects, a process known as post-processing \citep{Vietal13}.

As a result, understanding the role of the many different processes acting in clusters has been very challenging. For example, although both ram pressure stripping and harassment are expected to be much more effective in the cluster center, galaxies that pass through the cluster core may then subsequently move to much larger radii \citep{Voetal01}. In fact, this is likely typical as cosmological simulations show that the orbits of satellites on first infall into a cluster are highly eccentric orbits \citep{We2011}. This can resultantly smear out clear trends with clustocentric radius \citep{Smetal15}. However, when both clustocentric velocity and clustocentric radius are combined into a single plot known as a `phase-space diagram', some additional information can be gained over clustocentric distance alone.

Recently, there have been several studies connecting the star formation of galaxy populations to their location in phase-space diagrams. \cite{Maetal11} investigated the relationship between stellar mass and the star formation of SDSS galaxies with location in phase-space. Also, \cite{HFetal14} and \cite{Muetal14} studied the connection between star formation activity and location in phase-space for galaxies in galaxy clusters. In \cite{Gietal05}, a population of galaxies that have previously entered the cluster, but are now found beyond the virial radius, were identified (described as `back-splash' galaxies). In phase-space, back-splash galaxies were found to be located in a distinct region of phase-space, with a systematically lower velocity, than newly infalling galaxies. \cite{Ometal13} then demonstrated that galaxies at different positions in their orbit (first infaller, back-splash or virialized) occupy semi-distinct regions in projected phase-space diagrams. \cite{Haetal15} used phase-space to study the star formation quenching time scale of infalling galaxies in observed clusters, comparing with their simulation result. Similarly, \cite{Ometal16} combined phase-space diagrams of clusters in the SDSS to show that quenching of galaxies must occur quickly after a galaxy falls into a cluster, no longer than 1~Gyr after the first pericenter pass. 

Phase-space diagrams have also been used to study ram pressure stripping. In \cite{Jaetal15} a cosmological simulation of a cluster was combined with a semi-analytical recipe for gas stripping to show that the distribution of neutral gas detections in the cluster Abell~963 ($z$$\sim$$0.2$) could be well explained via a ram pressure stripping scenario \citep[see also][]{Jaetal16}.

Phase-space diagrams may also be useful for studying tidal mass loss in dense environments like groups and clusters. Using a model of the tidal field of cosmologically simulated cluster, \cite{Smetal15} showed that different regions of phase-space could be associated with different amounts of tidal mass loss. 

In this work, we attempt to extend on these pioneering studies by further developing how a galaxy's location in phase-space can provide valuable information on both time since infall and tidal mass loss. The simulations we use have excellent spatial and mass resolution, meaning they can be used to compare with observed galaxy samples down to stellar masses $\sim$$10^7$~M$_\odot$. The simulations include a full baryonic physics treatment, including stellar and AGN feedback. However, in this paper, we will focus on the evolution of the dark matter haloes only, as we have already considered the relative tidal stripping of dark matter and stars in detail in \citet{Smetal16}. In addition, by using fully hydrodynamical simulations, our simulations will naturally include baryonic feedback processes which can alter the profile of the dark matter haloes in which they reside \citep[e.g.,][]{Peetal16}, and thus could potentially alter their subsequent tidal mass loss. With these advanced hydrodynamical cosmological simulations, we quantify what the location of galaxies in phase-space can tell us about the the time since the galaxy fell into the cluster, and the amount of tidal mass loss they have suffered. We also quantify the variations that occur due to differing line-of-sights and cluster-to-cluster variations. We try to provide these results in an observer friendly format, and we also make the data in key figures available as supplementary data to try to encourage the use of phase-space diagrams as a tool for understanding environmental effects.

%%%%%%%%%%	 Table 1		%%%%%%%%%%%%%%%%%%%%%%%%%%%%%%%%%%%%%%%%%%%%%%%%%%%%%%%%%%%%%%%%%%%%%%%%%%%%%%%%%%%%%%%%%%%%%
\begin{table}[t] \begin{center}
  \caption{Information about 16 Clusters}
  \begin{tabular}{ccccc}
  \hline \hline

 \multicolumn{1}{c}{Name \tablenotemark{a}} & Mass (M\textsubscript{$\odot$}) \tablenotemark{b} & N\textsubscript{sub} \tablenotemark{c} & $R_{\rm{vir}}$~(Mpc) \tablenotemark{d} & $\sigma_{\rm{3D}}$~(km/s) \tablenotemark{e} \\
 \hline

C1 & 9.23 $\times$ 10$^{14}$ & 1231 & 2.54 & 1444 \\

C2 & 5.49 $\times$ 10$^{14}$ & 601 & 2.14 & 1154 \\

C3 & 2.30 $\times$ 10$^{14}$ & 485 & 1.60 & 920 \\

C4 & 2.26 $\times$ 10$^{14}$ & 342 & 1.59 & 880 \\

C5 & 2.25 $\times$ 10$^{14}$ & 326 & 1.59 & 881 \\

C6 & 1.97 $\times$ 10$^{14}$ & 357 & 1.52 & 816 \\

C7 & 1.91 $\times$ 10$^{14}$ & 258 & 1.50 & 900 \\ 

C8 & 1.70 $\times$ 10$^{14}$ & 265 & 1.45 & 787 \\

C9 & 1.56 $\times$ 10$^{14}$ & 251 & 1.41 & 724 \\

C10 & 1.47 $\times$ 10$^{14}$ & 276 & 1.38 & 720 \\

C11 & 7.03 $\times$ 10$^{13}$ & 139 & 1.08 & 624 \\

C12 & 6.76 $\times$ 10$^{13}$ & 101 & 1.06 & 561 \\

C13 & 6.32 $\times$ 10$^{13}$ & 101 & 1.04 & 524 \\

C14 & 5.77 $\times$ 10$^{13}$ & 76 & 1.01 & 481 \\

C15 & 5.30 $\times$ 10$^{13}$ & 84  & 0.98 & 470 \\

 \hline

 \end{tabular}
 \label{tab:all_clusters}
 \end{center}
 \bf{~~Notes.}
 \tablenotetext{1}{Name of the cluster.}
 \tablenotetext{2}{Virial mass of the cluster.}
 \tablenotetext{3}{The number of subhaloes within 3 virial radii at $z$$=$$0$ from the host cluster}
 \tablenotetext{4}{Virial radius of the cluster}
 \tablenotetext{5}{Velocity dispersion of subhaloes within the cluster}
\end{table}
%%%%%%%%%%	 end table1		 %%%%%%%%%%%%%%%%%%%%%%%%%%%%%%%%%%%%%%%%%%%%%%%%%%%%%%%%%%%%%%%%%%%%%%%%%%%%%%%%%%%%%%%%%%%%

%%%%%%%%%%%%%%%%%%%%%%%%%%%%%%%%%%%%%%%%%%%%%%%%%%%%%%%%%%%%%%%%%%%%%%%%%%%%%%%%%%%%%%%%%%%%%%%%%%%%%%%%%%%%%%%%%%%
%%%%%%%%%%%%%%%%%%%%%%%%%%%%%%%%%%%%%%%	Simulation		%%%%%%%%%%%%%%%%%%%%%%%%%%%%%%%%%%%%%%%%%%%%%%%%%%%%%%%%%%%%%%%%
%%%%%%%%%%%%%%%%%%%%%%%%%%%%%%%%%%%%%%%%%%%%%%%%%%%%%%%%%%%%%%%%%%%%%%%%%%%%%%%%%%%%%%%%%%%%%%%%%%%%%%%%%%%%%%%%%%%

\section[]{Simulation}
\label{sec:simulation}

%%%%%%%%%%%%%%%%%%%%%%%%%%%%%%%%%%%%%%%	Simulation-information		%%%%%%%%%%%%%%%%%%%%%%%%%%%%%%%%%%%%%%%%%%%%%%%%%%%%%%%%%%
\subsection{Cosmological Simulation}
\label{sec:simulation-information}

A more detailed description of the model is available in \cite{Chetal17} and \cite{Smetal16}; hence here we give only a brief summary. We ran cosmological hydrodynamic zoom-in simulations using the adaptive mesh refinement code RAMSES \citep{Te02}. We first run a large volume simulation for a cube of a side length of 200 Mpc/h using dark matter particles only adopting the WMAP7 cosmology \citep{Koetal11}: $\Omega_{\rm M}=0.272$, $\Omega_{\Lambda}=0.728$ and $H_{0}=70.4\, \rm km \, s^{-1} Mpc^{-1}$, $\sigma_{8}=0.809$, and $n=0.963$. We selected 15 high density regions in the cosmological volume and performed zoom-in simulations, this time including hydrodynamic recipes. The size of the zoom-in region is the maximum extent of all the particles that are within 3 viral radii of a cluster at $z$$=$$0$. For our typical clusters, this means roughly 20~Mpc comoving at $z$$=$$40$. The final resolution for force calculation is roughly 760~pc/h in length and $8 \times10^7\ \rm{M}_\odot$ in dark matter particle mass. We have adopted the baryon prescriptions of \cite{Duetal12} including gas cooling, star formation, stellar and AGN feedback. These  were applied in the Horizon-AGN simulation, and have been demonstrated to accurately reproduce the basic observational properties of galaxies such as the halo and stellar luminosity function, and their size and color \citep{Duetal14}.

%%%%%%%%%%%%%%%%%%%%%%%%%%%%%%%%%%%%%%%	Simulation-halo definition	%%%%%%%%%%%%%%%%%%%%%%%%%%%%%%%%%%%%%%%%%%%%%%%%%%%%%%%%%%

\subsection{Halo Definition}
\label{sec:simulation-halo}

After conducting zoom-in high resolution simulation of each cluster, we identify subhaloes in a cluster using the AdaptaHOP method \citep{Auetal04} in each snapshot from $z$$=$$3$ to $z$$=$$0$ and define physical quantities (e.g., virial mass, virial radius, radial vector, velocity vector) of each subhalo. The AdaptaHOP method identifies local maxima of the density field, and uses saddle points in the density field to differentiate between structures \citep{Twetal09}. Only structures with an average density larger than 178 times the average total matter density are identified as haloes. A subhalo is linked with the progenitor in the previous snapshot, and each snapshot is approximately 75~Myr apart, thus we have excellent time resolution which greatly aids the linking process. We choose to discard all subhaloes whose peak mass is lower than $3\times10^{10}\,\rm{M}_{\odot}$. This choice ensures that, even in the worst case, we can follow tidal mass loss of a halo down to 3$\%$ of its peak mass. It also ensures that, in almost all cases, our sample haloes have sufficient numbers of particles to avoid numerical effects resulting in artificial disruption \citep[e.g.,][]{Kletal99}.

We note that we do not directly consider the tidal mass loss of the stellar component of galaxies in these hydrodynamical simulations as we have previously studied this in detail in \citet{Smetal16}, and the limited spatial resolution of our simulations means that, in the lowest mass haloes, processes such as star formation, feedback, and tidal stripping are poorly resolved. However, with our choice of halo mass cut, almost all haloes contain a galaxy (93\%). This is important because it means that our haloes be considered as a proxy for a galaxy. The fact that most of our haloes contain galaxies also ensures that within our sample haloes baryonic feedback processes are in action, which can modify their dark matter profiles, and potentially impact their tidal stripping at later epochs. \citet{Peetal16} demonstrate, using very similar baryonic physics to our own, that above a halo mass of 1$\times$10$^{12}$~M$_\odot$, AGN feedback can significantly lower central halo densities. At this mass limit, all of our haloes contain galaxies. Thus it is reasonable that, in general, we use the term `galaxy' in place of `halo' throughout this paper. We confirm that the halo mass distribution follows a single power law down to this peak mass limit, as would be expected for well resolved haloes in $\Lambda$CDM. As our simulations are well resolved for haloes down to masses of 3$\times$10$^{10}$~M$_\odot$, our analysis can be applied to observed galaxies with stellar mass as low as $\sim$$10^{7}\,\rm{M}_{\odot}$, under the assumption of halo abundance matching \citep[see][]{Guetal10}. Because halo finders are not perfect, there is sometimes noisy halo mass evolution, from snapshot to snapshot. To avoid introducing this noise into our results, we smooth the mass evolution history of each subhalo, using a sigma-clipping technique. Table \ref{tab:all_clusters} contains a summary of information about the clusters and their subhaloes.

\subsection{Interlopers}
\label{sec:interlopersintro}
In observations of real clusters, objects that are actually non-members of the cluster may be projected by line-of-sight effects to appear close to the cluster. The same effect also occurs in phase-space, only the projection effect must occur in both the radius and velocity coordinate in order for a foreground or background galaxy to appear to lie within the projected phase-space diagram. These objects are described as `interlopers' \citep[e.g.,][]{Ometal16}.

In this work, our simulated clusters are zoomed out to 3 virial radii from their main halo. Therefore interlopers that come from beyond 3 virial radii would be missed. To account for this, we compliment our zoomed simulations with a single 200~Mpc/h box-side dark matter only cosmological simulation, which matches the resolution of our zoomed simulations down to our halo mass cut. We define interlopers as galaxies that fall within our projected phase-space diagrams ($0<R_{\rm proj}/R_{\rm vir}<3$ and $0<V_{\rm LOS}/\sigma_{\rm LOS}<3$), but are actually greater than 3 virial radii from the cluster. This is a little larger than the 2.5 virial radii definition used in \citet{Ometal16}, but this is necessary as our zoomed simulations are complete out to 3 virial radii. We will use this simulation to place contours on our projected phase-space diagrams, illustrating where few interlopers are found (see Section \ref{sec:result-master diagram} and \ref{sec:result-interloper}), or to quantify the number of interlopers in specified regions of phase-space (see Section \ref{sec:result-regional diagram}).

%%%%%%%%%%%%%%%%%%%%%%%%%%%%%%%%%%%%%%%%%%%%%%%%%%%%%%%%%%%%%%%%%%%%%%%%%%%%%%%%%%%%%%%%%%%%%%%%%%%%%%%%%%%%%%%%%%%
%%%%%%%%%%%%%%%%%%%%%%%%%%%%%%%%%%%%%%%	Result	%%%%%%%%%%%%%%%%%%%%%%%%%%%%%%%%%%%%%%%%%%%%%%%%%%%%%%%%%%%%%%%%%%%
%%%%%%%%%%%%%%%%%%%%%%%%%%%%%%%%%%%%%%%%%%%%%%%%%%%%%%%%%%%%%%%%%%%%%%%%%%%%%%%%%%%%%%%%%%%%%%%%%%%%%%%%%%%%%%%%%%%

\section[]{Results}
\label{sec:result}

%%%%%%%%%%%%%%%%%%%%%%%%%%%%%%%%%%%%%%%	Result-Phase space Diagram	%%%%%%%%%%%%%%%%%%%%%%%%%%%%%%%%%%%%%%%%%%%%%%%%%%%%%%
\subsection{Typical orbital trajectories in Phase-space}
\label{sec:result-psd}

The phase-space diagram we use involves plotting velocity on the y-axis, versus radius on the x-axis. In the simulations, the orbital velocity and clustocentric radius of the galaxy are used, measured with respect to the cluster frame-of-reference. The radius axis is normalized by the cluster virial radius $R_{\rm vir}$. Meanwhile, the velocity axis is normalized by the cluster velocity dispersion $\sigma_{\rm 3D}$, measured for all haloes within $R_{\rm vir}$.  

\begin{figure}																			%
\centering 																			%
\includegraphics[width=0.45\textwidth]{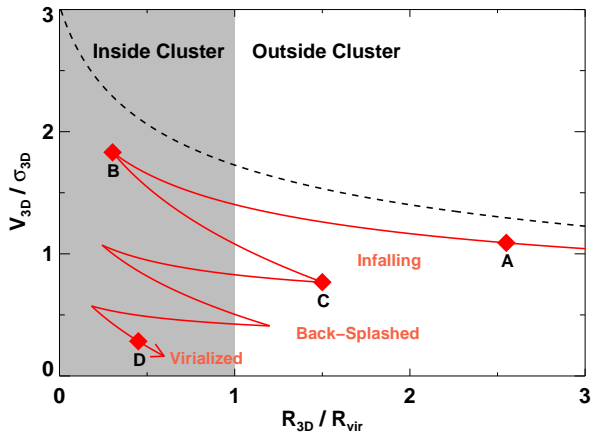}												%
\caption{Toy schematic of a galaxy's trajectory after inalling into the cluster, shown using normalized orbital velocity, and normalized clustocentric radius. The red solid line is the trajectory of the galaxy up until $z=0$. The black-dashed line is the escape velocity curve of the cluster, calculated as described in the text. Initially the galaxy is found in the `infalling' region (between A and B). After first pericenter, it approaches apocenter at C. If the apocenter is beyond the cluster virial radius, then the galaxy is found in the `back-splash' region. Between C and D, the object settles into the lower-left hand corner of the diagram, known as the `virialized region'.}									%
\label{fig:fig cartoon psd}																	%
\end{figure}

A schematized example of such a phase-space diagram is shown in Figure \ref{fig:fig cartoon psd}. The black dashed line is the escape velocity of the cluster at that radius. This is derived assuming an NFW halo density profile \citep[]{NFW96} and assuming a concentration parameter $c=6$, which is a typical value for cluster mass NFW haloes \citep{Gietal04b}. For an NFW halo, the escape velocity is
\begin{eqnarray}
v_{\rm esc}=\sqrt{\frac{2GM_{\rm vir}}{R_{\rm vir}}K(s)}\rm{,}
\end{eqnarray}
where
\begin{equation}
K(s)=g_c\,\frac{\ln(1+cs)}{s}\rm{,}
\end{equation}
\begin{equation}
s=\frac{R_{\rm 3D}}{R_{\rm vir}}\rm{,}
\end{equation}
and
\begin{equation}
g_c=\left[\ln(1+c)-\frac{c}{1+c}\right]^{-1}\rm{.}
\end{equation}

%%%%%%%%%%%%%%%%%%%%%%%%%%%%%%%%%%%%%%%%%%%%%%%%%%%%%%%%%%%%%%%%%%%%
%%%%%%%%%		Figure 1		%%%%%%%%%%%%%%%%%%%%%%%%%%%%%%%%%%%%%%%%%%%%%%
%%%%%%%%%		End Figure 	%%%%%%%%%%%%%%%%%%%%%%%%%%%%%%%%%%%%%%%%%%%%%%
%%%%%%%%%%%%%%%%%%%%%%%%%%%%%%%%%%%%%%%%%%%%%%%%%%%%%%%%%%%%%%%%%%%%

%%%%%%%%%%%%%%%%%%%%%%%%%%%%%%%%%%%%%%%%%%%%%%%%%%%%%%%%%%%%%%%%%%%%
%%%%%%%%%		Figure 2		%%%%%%%%%%%%%%%%%%%%%%%%%%%%%%%%%%%%%%%%%%%%%%
\begin{figure*}																			%
\centering 																			%
\includegraphics[width=0.95\textwidth]{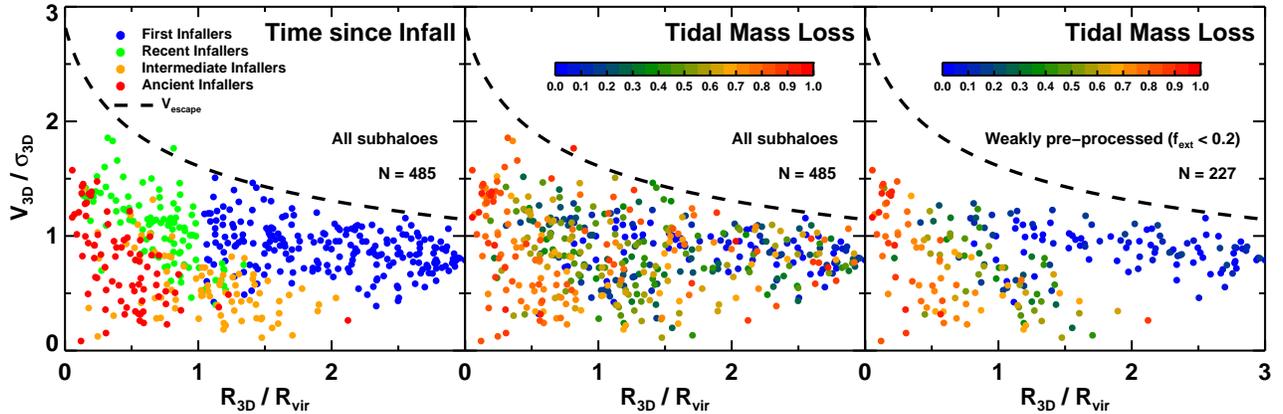}												%
\caption{Phase-space diagrams of a single massive cluster, Cluster C3 ($\sim2.3\times10^{14}$M$_{\odot}$). The black-dashed line is the escape velocity curve. In the left panel, symbol color indicates time since the galaxy fell into the cluster, binned into 4 time bins (First Infallers, Recent Infallers, Intermediate Infallers and Ancient Infallers, for detail see text), where color indicates the time bin (see legend). Galaxies in each time bin tend to occupy fairly distinct regions in the phase-space diagram. In the middle and right panel, color indicates the fraction of the total halo mass that has been tidally stripped. Blue colors indicate weak tidal mass loss, and red colors indicate very strong tidal mass loss (see color bar). The left and middle panel include all the galaxies of the cluster. The right panel contains a subsample of the galaxies, which have suffered low mass loss external to the cluster ($f_{\rm ext}$$<$$0.2$). With the exclusion of pre-processed galaxies, the plot of time since infall (left panel) and the tidal mass loss (right panel) share a resemblance.}	
\label{fig:fig 3d psd}																		%
\end{figure*}																			%
%%%%%%%%%		End Figure 	%%%%%%%%%%%%%%%%%%%%%%%%%%%%%%%%%%%%%%%%%%%%%%
%%%%%%%%%%%%%%%%%%%%%%%%%%%%%%%%%%%%%%%%%%%%%%%%%%%%%%%%%%%%%%%%%%%%

The red line shows a fairly typical trajectory that a galaxy that falls into the cluster might take. Between A and B, the galaxy falls into the cluster for the first time, tending to remain quite close to the escape velocity curve. The galaxy reaches pericenter at B. Between B and C, the galaxy approaches its first apocenter. At C, the galaxy reaches apocenter, and as it is currently found beyond the cluster virial radius, it is technically a `back-splash' galaxy. Between C and D, the galaxy orbits within the cluster, settling into the lower-left hand corner of the figure, known as the virialized region. Previous studies \citep[see e.g.,][]{Be85,Ometal13} have shown that such an orbital path is typical. The settling of galaxies into the lower-left hand corner of the phase-space diagram is driven by several mechanisms. For massive galaxies, dynamical friction can play a role, dragging galaxies closer to the cluster center. However, for lower mass galaxies, the large mass ratio difference between the galaxy and the cluster means dynamical friction is rather ineffective. Therefore, some of the orbital change may be driven by the changing shape of the potential well in which the galaxies are orbiting, as the cluster grows in mass by accretion \citep{Gietal04a}. In addition, as the cluster grows in mass, it also grows in virial radius. Therefore galaxies which fell in a long time ago, and are now found in the virialized region would initially have been orbiting in a cluster with a smaller virial radius than today.

We note that some galaxies may have already suffered environmental effects before falling into the cluster (e.g., at A on Figure \ref{fig:fig cartoon psd}). As a result, we will deliberately separate out strongly pre-processed galaxies when studying tidal mass loss in phase-space (see Section \ref{sec:result tml}).

One of the main goals of this paper is to show how phase-space diagrams can be applied to observed clusters, in order to better understand environmental effects. For this purpose we also produce `projected' phase-space diagrams. In these diagrams, the orbital velocity on the y-axis becomes a line-of-sight velocity, and the clustocentric radius becomes a projected distance from the cluster center, thereby using directly observable quantities. As a result, there is some mixing of galaxy locations in phase-space due to projection effects, but projected phase-space diagrams have a significant advantage in that they can be compared directly with observed clusters. We will present our result first on the 3D analysis and then the projected phase-space analysis.

\subsection{A representative example of phase-space diagrams for the massive cluster C3}
\subsubsection{Time since infall}
To begin with, we measure the time since a subhalo first enters the cluster virial radius, and this time is referred to as the `time since infall'. Estimating the time since infall using a phase-space diagram could be useful for understanding environmental effects, as different environmental mechanisms might be expected to act on differing timescales. For example, ram pressure is often described as acting very rapidly, whereas starvation is believed to occur on much longer timescales \citep{Boetal08}. Also, as illustrated in Figure \ref{fig:fig cartoon psd}, if galaxies tend to follow the typical trajectory indicated, then different locations in phase-space can be expected to correlate with particular times since infall. 

For illustrative purposes, in Figure \ref{fig:fig 3d psd} we first consider all of the galaxies in a single fairly massive cluster, cluster C3 ($\sim$$2.30\times10^{14}\,\rm{M}_{\odot}$), that can be considered a representative example. We will first focus just on the left panel, and return to the middle and right panel in the following section. In the left panel, each subhalo at $z$$=$$0$ is shown by a data point, and the color of the data points indicates the time since infall into the cluster. Blue symbols have not yet entered the cluster and are infalling towards the virial radius for the first time, so are referred to as `first infallers'. Green symbols are referred to as `recent infallers' as they have recently fallen into the cluster in the last 0 - 3.63~Gyr. Yellow symbols fell into the cluster between 3.63 and 6.45~Gyr ago, and are referred to as `intermediate infallers'. Finally red symbols are referred to as `ancient infallers' as they fell into the cluster more than 6.45~Gyr ago (see the Table \ref{tab:tsf groups}). These time bins are chosen to ensure equal number statistics in each time bin, for all the clusters combined.

\begin{table}[t]
 \begin{center}
  \caption{Criteria of time since infall groups}
  \begin{tabular}{cc}
  \hline \hline

  \multicolumn{1}{c}{Groups} & Time bin(Gyr) \\
 \hline
First Infallers & Not fallen yet \\
Recent Infallers & $0<t_{\rm inf}<3.63$ \\
Intermediate Infallers  & $3.63<t_{\rm inf}<6.45$ \\
Ancient Infallers & $6.45<t_{\rm inf}<13.7$ \\
 \hline
 \end{tabular}
 \label{tab:tsf groups}
 \end{center}
\end{table}

The origin of this separation is best understand by comparison with the typical trajectory in Figure \ref{fig:fig cartoon psd}. Galaxies that are infalling for the first time (blue) tend to stay close to the escape velocity curve, as they fall within the cluster's potential, moving to smaller radius while simultaneously gaining velocity. Once they enter the cluster virial radius, the symbols become green and they reach peak velocity at their smallest radius (i.e. at pericenter). Hence recent infallers tend to be found at small radius and high velocity. After passing pericenter, galaxies move towards apocenter and for some this is beyond the virial radius of the cluster in the back-splash region (see yellow points beyond 1 virial radius). Where a three dimensional velocity and radius are used, it is clear that these `back-splash' galaxies are quite well separated from the blue first infallers data points in a three dimensional phase-space diagram\footnote{Although they are not so cleanly separated in {\it{projected}} phase-space diagrams \citep{Ometal13}.}, as they have a systematically smaller velocity. Following first apocenter, the galaxy may pass several more pericenters and apocenters while settling into the bottom left hand corner of the phase-space diagram, known as the virialized region which is dominated by ancient infallers (red points).

As a result of galaxies having several orbits since first apocenter, there is a small amount of mixing of the orange points, with the red and the green points. However, fortunately galaxies with eccentric orbits spend the majority of their time near apocenter. As a result, even though intermediate infallers may spend some time at a wide range of radii and velocity, they tend to spend most of their time at large radius and small velocity. Thus the mixing of the orange points with the red and green points is not very severe.

This quite clean separation of galaxies with different times since infall in three dimensional phase-space diagrams has been shown previously \citep[e.g.,][]{Ometal13,Maetal11}. Here, we independently confirm their general trends with our group/cluster samples. We note that previous studies \citep[e.g.,][]{Ometal13} have demonstrated that by including a positive or negative sign on $V_{\rm 3D}$, for infalling or outward moving galaxies respectively, the separation between different time since infall regions can be increased. However, in this study, we choose not to include the positive or negative sign because, in this way, in order to move from our 3D phase space diagram to its projected equivalent, it is only necessary to take a single projection of the 3D results. Thus, it is easier to intuitively see how our 3D phase-space diagrams are related to our projected phase-space diagrams. In the following sections, we also try to take our analysis one step further by considering the link between time since infall and tidal mass loss. We will study this link in three dimensional phase space diagrams (Section \ref{sec:result tml}), and also in projected phase-space diagrams (Section \ref{sec:result-projected psd}) that can be compared directly with observed populations of galaxies in groups and clusters.

%%%%%%%%%%%%%%%%%%%%%%%%%%%%%%%%%%%%%%%%%%%%%%%%%%%%%%%%%%%%%%%%%%%%
%%%%%%%%%		Figure 3		%%%%%%%%%%%%%%%%%%%%%%%%%%%%%%%%%%%%%%%%%%%%%%
\begin{figure}																			%
\centering 																			%
\includegraphics[width=0.48\textwidth]{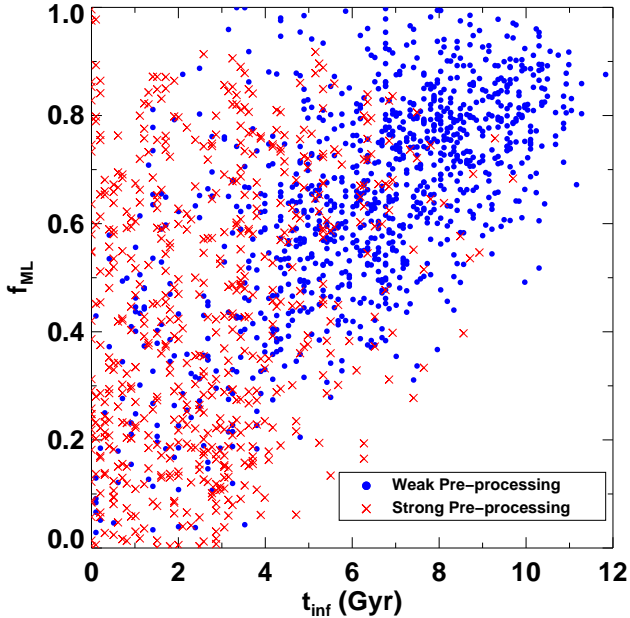}													%
\caption{The correlation between fractional tidal mass  loss $f_{\rm ML}$ and time since infall into the cluster $t_{\rm inf}$. Data points are galaxies from all of the clusters in our sample. We make two subsamples based on the strength of pre-processing they experience. The weakly pre-processed sample lose less than 20$\%$ of their total mass loss before entering the cluster (dark blue filled circles). The strongly pre-processed subsample lose more than 80$\%$ of their total mass loss before entering the cluster (red crosses). In the absence of pre-processing effects, a clear linear trend between time since infall and mass loss is seen.}																			%
\label{fig:fig relation}																		%
\end{figure}																			%
%%%%%%%%%		End Figure 	%%%%%%%%%%%%%%%%%%%%%%%%%%%%%%%%%%%%%%%%%%%%%%
%%%%%%%%%%%%%%%%%%%%%%%%%%%%%%%%%%%%%%%%%%%%%%%%%%%%%%%%%%%%%%%%%%%%

\subsubsection{Tidal mass loss}
\label{sec:result tml}

In the middle panel of Figure \ref{fig:fig 3d psd}, the symbols are instead colored by the total fractional tidal mass loss suffered by the dark matter haloes of each galaxy. To calculate the fractional tidal mass loss, we consider the final halo mass of a galaxy at $z$$=$$0$, and compare to its peak mass in the past. In this way, we will be sensitive to any mass loss a halo suffers since it reached its peak mass although, as we will show, some of this mass loss may occur outside of the cluster environment. The fractional tidal mass loss $f_{\rm ML}$ is formulated as follows:
\begin{equation}
\label{fracmasslosseqn}
f_{\rm ML}=1-\frac{M_{\rm now}}{M_{\rm peak}}
\end{equation}
where $M_{\rm now}$ is the final mass (at $z$$=$$0$) and $M_{\rm peak}$ is the maximum mass in the mass evolution history of a subhalo.

Comparing the left and middle panel of Figure \ref{fig:fig 3d psd}, it is clear that there are some similarities. The blue points in the left panel (first infallers) tend to lie in approximately the same region in phase-space as the blue points in the middle panel (weak tidal mass loss). Similarly the red points in the left panel (ancient infallers), tend to lie in approximately the same region as the red points in the right panel (strong tidal mass loss). {\it{This implies a trend for increasing tidal mass loss with increasing time spent in the cluster.}} However, there is clearly some considerable scatter in that trend, that causes the separation not to be as neat as was seen in the left panel. For example, those galaxies that have never even entered the cluster (blue points, left panel) have in some cases suffered quite strong tidal mass loss (some red and orange points are mixed in with the blue points).

One of the key reasons for this mixing is due to our definition of the fractional tidal mass loss in Equation \ref{fracmasslosseqn}. Because we use the peak mass, not all the halo mass loss may be due to the cluster tides -- some may have occurred external to the cluster, for example due to tides in groups. To quantify this, we instead split $f_{\rm ML}$ into two summed components:

\begin{equation}
f_{\rm ML}=f_{\rm int}+f_{\rm ext}
\end{equation}
where $f_{\rm ext}$ is the fractional tidal mass loss that occurred before entering the cluster virial radius (i.e. externally), and $f_{\rm int}$ is the fractional tidal mass loss that occurred after entering the cluster (i.e. internally).

%%%%%%%%%%%%%%%%%%%%%%%%%%%%%%%%%%%%%%%%%%%%%%%%%%%%%%%%%%%%%%%%%%%%
%%%%%%%%%		Figure 4		%%%%%%%%%%%%%%%%%%%%%%%%%%%%%%%%%%%%%%%%%%%%%%
\begin{figure}																			%
\centering 																			%
\includegraphics[width=0.48\textwidth]{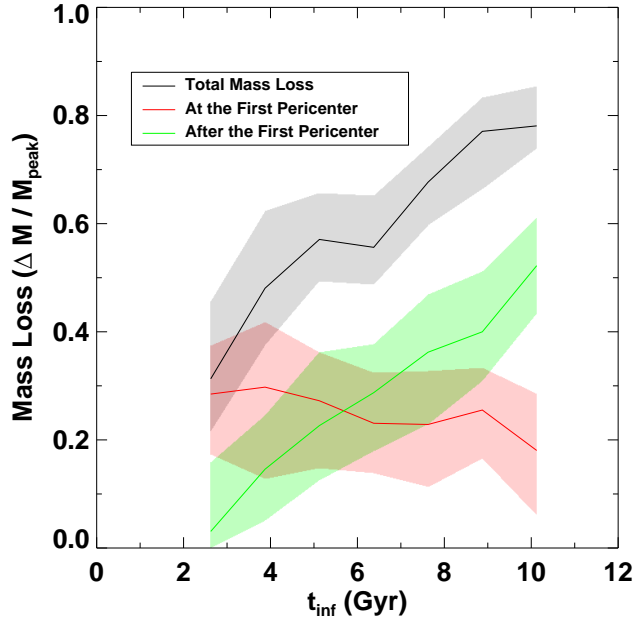}													%
\caption{For the weak pre-processing sample (blue circles in Figure \ref{fig:fig relation}), we measure the median fraction of a galaxy's mass that is lost, as a function of time since infall (black line). We separate this total mass loss into median mass lost by first pericenter (red line), and median mass lost since the first pericenter (green line). The shading indicates the first and third quartile of the data points in each time bin. We truncate the fits at t$_{\rm{inf}}<2.75~$Gyr, as there are very few galaxies which fell in so recently and reached pericenter.}																			%
\label{fig:fig peritest}																		%
\end{figure}																			%
%%%%%%%%%		End Figure 	%%%%%%%%%%%%%%%%%%%%%%%%%%%%%%%%%%%%%%%%%%%%%%
%%%%%%%%%%%%%%%%%%%%%%%%%%%%%%%%%%%%%%%%%%%%%%%%%%%%%%%%%%%%%%%%%%%%

To isolate the effect of external mass loss on the mixing, we choose a subset of haloes which suffered weak mass loss outside the cluster ($f_{\rm ext}$$<$$0.2$). The result is shown in the right panel of Figure \ref{fig:fig 3d psd}. There is clearly a close resemblance in the distribution of galaxies with a particular time of infall (left panel), and a particular amount of tidal mass loss (right panel), once we have excluded cases where strong mass loss occurs exterior to the cluster.

Therefore, we can already deduce that there is a direct correlation between time since infall and tidal mass loss, with the pre-processing causing spread from that relation. This correlation is shown in Figure \ref{fig:fig relation}, whose data are drawn from our complete set of clusters. The dark blue filled circles show a sub-sample of weakly pre-processed galaxies, which have lost less than 20$\%$ of their total mass loss before entering the cluster. The red crosses show a sub-sample of strongly pre-processed galaxies, which have lost more than 80$\%$ of their total mass loss before entering the cluster.

As deduced, the direct correlation between time since infall and tidal mass loss is very clear, once we exclude strongly preprocessed galaxies. Interestingly, it is also very linear with time. Meanwhile, the heavily pre-processed galaxies (red crosses) can be seen to spread vertically upwards from the linear trend to higher mass losses. The distance they spread from the linear trend is small when infall occurred a long time ago, and much larger for recent infallers. This means the effects of pre-processing are much more significant for galaxies falling into the cluster now than it was in the past. In fact, if we return to Figure \ref{fig:fig 3d psd}, it can be seen (comparing the middle and right panel) that the scatter in the $f_{\rm ML}$ distributions due to pre-processing is not very strong for the intermediate or ancient infaller populations, and is most significant for the first infaller populations. To try to better quantify the effects of tidal mass loss from pre-processing for the all of the simulated clusters in our sample, we calculate the ratio $f_{\rm ext}$/$f_{\rm ML}$ for all galaxies within one virial radius of a cluster. This ratio is essentially a measure of how much of the total mass loss occurred outside the cluster. For the total sample, the median value is only 0.27, meaning that typically the cluster tidal mass loss dominates significantly over pre-processing for the cluster population. In fact, the median value of $f_{\rm ext}$/$f_{\rm ML}$ reduces to only 0.15 if we restrict our sample to galaxies with high times since infall ($>$ 6 Gyr). This further confirms that the significance of pre-processing for the cluster population is much lower for galaxies with earlier times since infall. The physical reason behind this is likely that, just like in the cluster, the effects of tidal stripping in groups also increases with time. Thus those galaxies that are falling into the cluster now have had the maximum time for pre-processing to be effective.

In the following, we try to better understand the linear trend with time since infall shown by the weakly pre-processed in Figure \ref{fig:fig relation}, and sources of scatter in that trend. In the simplest scenario, galaxies would be truncated to their tidal radius by their first pericentric passage. In this case, if the galaxy does not radially decay significantly, then spending additional time in the cluster would not result in additional mass loss. This is contrary to the clear trend we see with time spent in the cluster. We quantify how much of our galaxies total mass at infall is lost by the first pericenter passage, and how much is lost subsequently. The results are shown in Figure \ref{fig:fig peritest}. In the figure, the solid lines are median values of the sample. In fact, the median amount of mass lost at first pericenter (red line) is about 20-30$\%$, and roughly constant with time. We also separately confirm that we do not see a strong evolution in the median pericentric distances with time. Therefore, the first pericentric passage cannot contribute to the linear trend seen by the weakly pre-processed systems (black line). Instead, it is the amount of mass loss subsequent to pericentric passage (green line) that is a clear function of time spent in the cluster, and that is fully responsible for the linear trend with time since infall. We note that, for objects which infall less than 6~Gyr ago, the mass loss at first pericentric passage is, on average, the dominant source of mass loss (the red line is above the green line). For objects that infall more than 6 Gyr ago, mass loss from high speed encounters with cluster members, coupled with subsequent pericentric passages \citep[e.g.,][]{TB04}, becomes the dominant source of mass loss. Spread in the linear trend of the weakly pre-processed sample (grey shading) may be partly a result of the variation in pericenter distances (e.g., red shading), combined with the varying halo concentration of galaxies undergoing tidal stripping \citep{Smetal13}.
%%%%%%%%%%%%%%%%%%%%%%%%%%%%%%%%%%%%%%%%%%%%%%%%%%%%%%%%%%%%%%%%%%%%
%%%%%%%%%		Figure 5		%%%%%%%%%%%%%%%%%%%%%%%%%%%%%%%%%%%%%%%%%%%%%%
\begin{figure*}																			%
\centering 																			%
\includegraphics[width=0.938\textwidth]{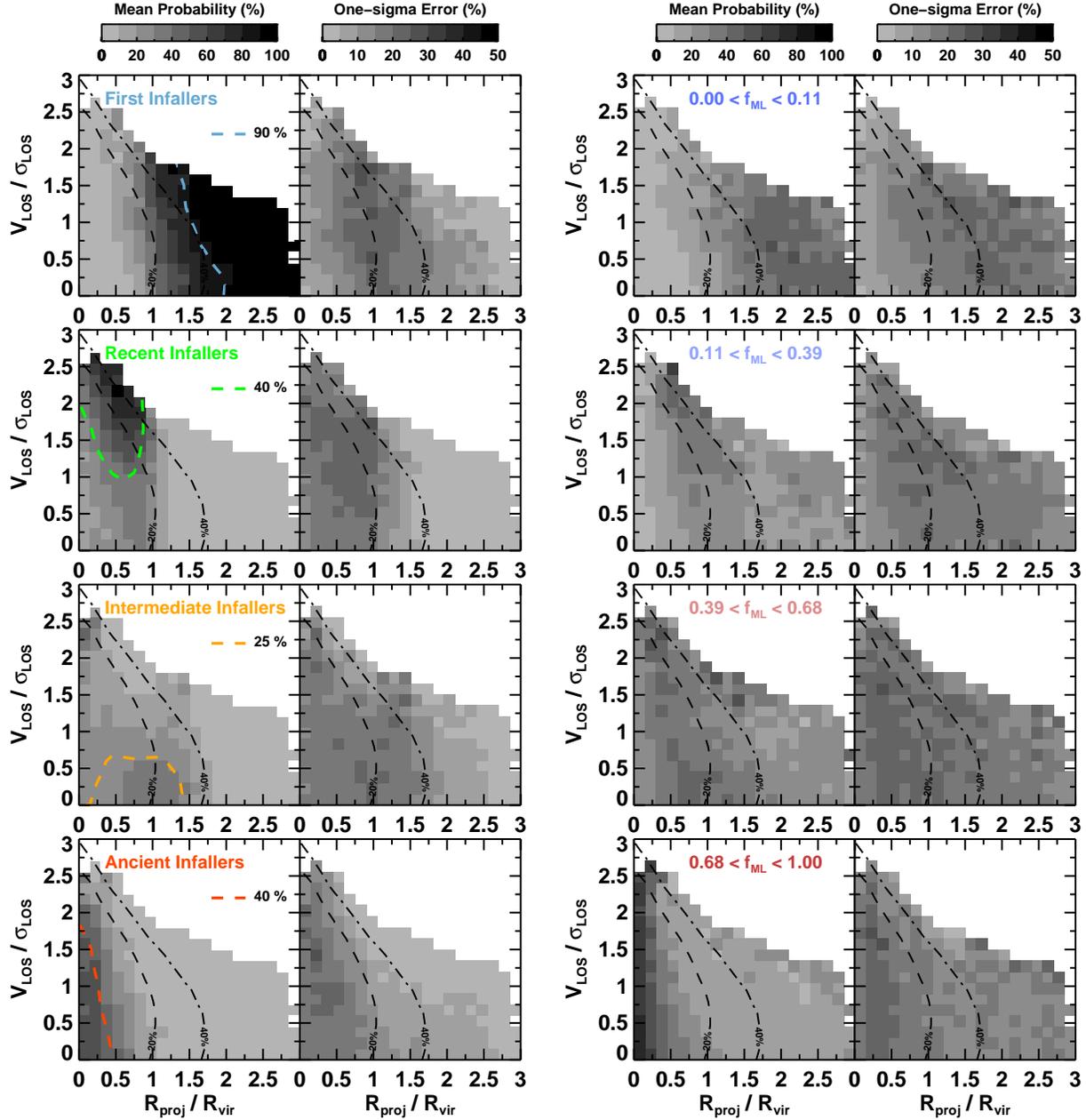}												 	%
\caption{Projected phase-space diagrams, using observable quantities such as line-of-sight velocity and projected radius, so as they are easily comparable with observations, once accounting for interlopers. In these master diagrams, all clusters are combined into a single figure (see text for details). In the first column, the grey scale shows the mean probability of an object having fallen into the cluster at a specified time (e.g., top to bottom: first infallers, recent infallers, intermediate infallers or ancient infallers). In the second column, we show the one-sigma errors in probabilities that arise due to differing lines-of-sight and cluster-to-cluster variations. The third and fourth column are similar to the first and second column, only the grey scale refers to tidal mass loss, instead of time since infall. Colored-dashed lines in the first column are isocontours of constant probability, with arbitrary values (indicated in the legend). We do not include interlopers in the calculation of the mean probability or one-sigma error. Instead, we indicate the presence of interlopers with the black dashed and black dot-dashed lines, which are contours showing where the interloper fraction is 20$\%$ and 40$\%$, respectively (see Section \ref{sec:result-interloper} for details). Each time since infall population has a different peak probability (i.e. a different maximum value of the pixels in that panel), therefore the isocontour values are chosen so as each contour best highlights the region of peak probability of that infall population, and also to avoid the contours overlapping if shown together on one plot (as will be the case in later figures).}																			%
\label{fig:fig mas psd}																	%
\end{figure*}																			%
%%%%%%%%%		End Figure 	%%%%%%%%%%%%%%%%%%%%%%%%%%%%%%%%%%%%%%%%%%%%%%
%%%%%%%%%%%%%%%%%%%%%%%%%%%%%%%%%%%%%%%%%%%%%%%%%%%%%%%%%%%%%%%%%%%%

%%%%%%%%%%%%%%%%%%%%%%%%%%%%%%%%%%%%%%%	Result-Projected Phase space Diagram	%%%%%%%%%%%%%%%%%%%%%%%%%%%%%%%%%%%%%%%%%%%%%%%%

\subsection{Projected Phase-space Diagrams}
\label{sec:result-projected psd}

In the previous subsection we have seen that if we use the three-dimensional orbital velocity and radius, then location in phase-space can provide valuable information about cluster galaxies time since infall, and even tidal mass loss due to the cluster potential.

However, the figures we have shown so far have limited applicability to sample of real, observed galaxies. This is because, in almost all cases, it is very challenging to determine the true three-dimensional radius and/or orbital velocity of a real galaxy in a cluster. Typically, the observed distances will be {\it{projected}} distances ($R_{\rm proj}$), and the observed velocities will be {\it{line-of-sight}} velocities ($V_{\rm LOS}$). To make phase-space diagrams from our cosmological simulations that are directly comparable with samples of observed cluster galaxies, we must first project orbital velocities and clustocentric radii down a single line of sight and sky plane, respectively. We consider multiple, randomly chosen lines-of-sight to each cluster. Each galaxy changes its $V_{\rm LOS}$ and/or $R_{\rm proj}$, depending on the particular line-of-sight that we look along. In case of $V_{\rm LOS}$, the Hubble flow correction is included as follow, assuming a line-of-sight along the z-axis:
\begin{equation}
\label{interloper}
V_{\rm LOS}\,=\,|v_{\rm z}\,+\,H_{\rm 0}\,\times\,r_{\rm z}|
\end{equation}
where $v_{z}$ and $r_{z}$ are the z-components of clustocentric velocity and radial vectors, respectively.

For the projected phase-space diagrams, we normalize the resulting projected radius axis by the virial radius of the cluster $R_{\rm vir}$, as previously. However, for the projected velocity axis, we normalize by the dispersion in galaxy line-of-sight velocities $\sigma_{\rm LOS}$, which is measured down the projected line-of-sight for each rotation of the cluster, much like in observed clusters. For a large sample of galaxies, $\sigma_{\rm LOS}\sim\sigma_{\rm 3D}/\sqrt{3}$. 

By normalizing the axes in this manner, we can place the galaxies from all 15 groups and clusters on a single, master diagram of projected phase-space. However, each group or cluster has differing number of galaxies. Therefore we first calculate a projected phase-space diagram for each cluster individually, considering multiple lines-of-sight (we arbitrarily choose 1000 line-of-sights, but we find negligible differences if we increase the number of lines-of-sight by a factor of five). We then combine all the cluster phase-space diagrams together in such a way that each has equal weight in the final master diagram. In this manner, we ensure that our master diagrams are not biased towards the clusters with the most galaxy members (i.e. the more massive clusters). Also, as we will demonstrate in Section \ref{sec:result-psd dependency-cluster}, our projected phase-space diagrams do not depend strongly on the mass of the group or cluster.

\subsubsection{Probability of time since infall in a projected phase-space diagram}
\label{sec:result-master diagram}

The first column of Figure \ref{fig:fig mas psd} shows where galaxies with different times since infall (from top to bottom: first infallers, recent infallers, intermediate infallers, and ancient infallers) tend to be found in projected phase-space. {\it{The grey scale indicates the probability that a galaxy at that location fell into the cluster at the specified time since infall.}} The second column gives the one-sigma error  on that probability, based on the standard deviation of variations induced by differing line-of-sights combined with cluster-to-cluster variations.

To produce the probability and standard deviation in a particular pixel, we first consider an individual cluster. We choose a single line-of-sight through the cluster. We then measure the fraction of galaxies falling in that pixel that have the specified time since infall compared to the total number of galaxies that fall in that pixel. We then randomly change our line-of-sight, and re-calculate the fraction in that pixel. With each spin of the cluster, we build up a series of fractions for that pixel, and we can calculate a mean and standard deviation of the series. Clearly we can conduct the same process in parallel for all the pixels to build up a probability map (and associated one-sigma errors due to differing lines-of-sight) over all of the phase-space for the individual cluster. However, to avoid noise due to low number statistics, we exclude fractions from our mean (and standard deviation) if the total number of galaxies falling in the pixel were less than four, for that particular line-of-sight. 

We then repeat this process for all of the groups and clusters, so as we have a probability map (and associated one-sigma errors) for each system. In a final step, we combine all of the individual maps into a single master map. In each pixel of the master map, the mean value of the probability of all the individual groups/clusters is calculated. In this way, each group/cluster provides an equal weight to the master maps, so as we are not biased towards more massive clusters which have greater numbers of satellites. If a pixel in the master map does not contain three or more clusters, it is excluded so as to again avoid low number statistics. This exclusion has a negligible impact on our results, except at the very upper boundary of the master map, where few objects fall. In a similar way, the standard deviation of the master map is calculated from the standard deviations of all the individual groups/cluster using standard error propagation formulae. The resulting standard deviations contain variations due to differing line-of-sights, combined with cluster-to-cluster variations. 

In this way we can build a probability (and one-sigma error map) for each population with a different time since infall (e.g., first infallers, ancient infallers, etc), as shown in the first and second column of Figure \ref{fig:fig mas psd}. It is interesting to note that, despite the smearing of galaxy positions due to projection effects, galaxies with different times since infall still tend to have their peak probabilities in distinct regions of phase-space, in a similar way as was seen for Cluster C3 when using a non-projected phase-space plot (see the left panel of Figure \ref{fig:fig 3d psd}). However,  it is clear that differences down each line-of-sight must contribute to the one-sigma errors provided in the second and fourth column of Figure \ref{fig:fig mas psd}. We note that we do not include interlopers when calculating the mean probability and standard deviation. Instead we indicate where interlopers are found using the black dashed and black dot-dashed contours which show where the interloper fraction is 20$\%$ and 40$\%$ of the total galaxies in that location in phase-space.

%%%%%%%%%%%%%%%%%%%%%%%%%%%%%%%%%%%%%%%%%%%%%%%%%%%%%%%%%%%%%%%%%%%%
%%%%%%%%%		Figure 6		%%%%%%%%%%%%%%%%%%%%%%%%%%%%%%%%%%%%%%%%%%%%%%
\begin{figure*}																			%
\centering 																			%
\includegraphics[width=0.85\textwidth]{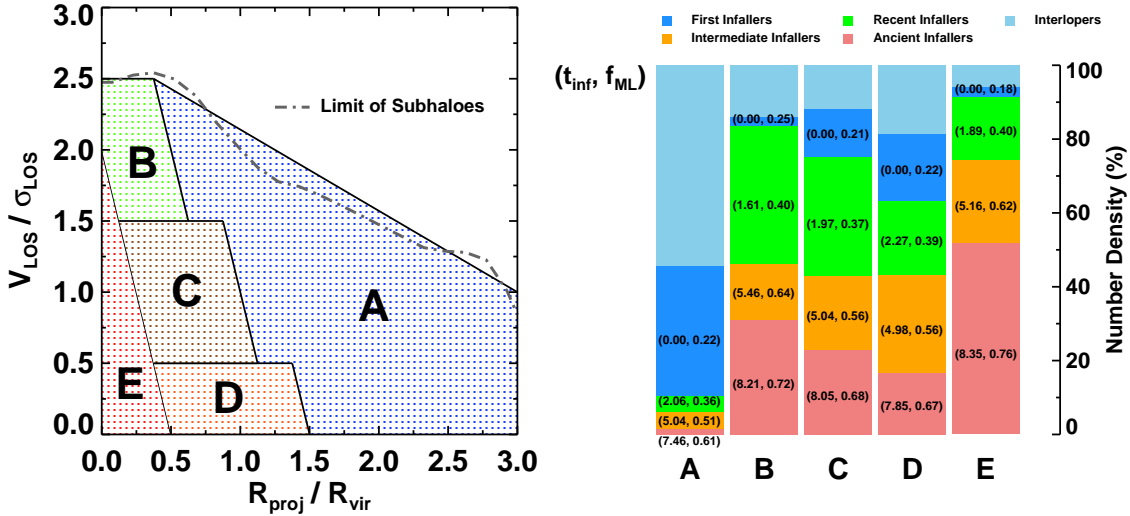}													%
\caption{Left panel: A projected phase-space diagram, split up into distinct regions (labelled A-E). Regions are chosen to try to maximize the fraction of a particular time since infall population. The grey-dashed line indicates the limit of subhaloes after 1000 random rotations. Right panel: Information about each region. In each bar graph, colors indicate the corresponding time since infall group (see legend), and vertical height of the color bar indicates the fraction of the total galaxies in that region. The numbers in brackets indicate the mean time since infall, and mean tidal mass loss fraction in that bar, respectively.} 																			%
\label{fig:fig region psd}																	%
\end{figure*}																			%
%%%%%%%%%		End Figure 	%%%%%%%%%%%%%%%%%%%%%%%%%%%%%%%%%%%%%%%%%%%%%%
%%%%%%%%%%%%%%%%%%%%%%%%%%%%%%%%%%%%%%%%%%%%%%%%%%%%%%%%%%%%%%%%%%%%

We note that, if the crossing-time of the cluster was a strong function of cluster mass, then combining clusters with different masses into the master diagrams could potentially result in smearing out of the regions where galaxies with a particular time since infall have their highest probabilities to be found. However, from the first column of Figure \ref{fig:fig mas psd}, it is clear that the regions are not strongly smeared out. We also measure crossing-time of subhaloes for each cluster, which should be related to the time interval between the first pericenter to the first apocenter for galaxies that fell in the most recently. We find that this time interval is 1.2~Gyr with a standard deviation of 0.5~Gyr for all clusters, with no indication of a trend with cluster mass. This is probably partly due to the fact that the free fall time is directly related to the density of a system, and each halo has a mean density which is $\sim$200 times the critical density of the Universe. This means binning our galaxies into the same bins of time since infall for all our groups/clusters is physically reasonable. We additionally note that we consider the dependencies of our phase-space diagrams on cluster mass (and galaxy mass) in Section \ref{sec:result-psd dependency}, and find only weak dependencies. The results we have described so far independently confirm the behavior described by previous authors (e.g., \citealp{Ometal13}, \citealp{Haetal15}) between time since infall and location in phase-space diagrams, but now using hydrodynamical cosmological simulations, and with excellent mass resolution. In addition, beyond reproducing the same basic trends, we attempt to better quantify what location in phase-space can tell us about time since infall. Our panels in Figure \ref{fig:fig mas psd} are designed to allow the mean probability (and the one-sigma error in that quantity) to be read off from the figure. We include two contours (black dashed is 20$\%$ and black dot-dashed is 40$\%$) to illustrate the presence of interlopers (for their definition, see Section \ref{sec:interlopersintro}). We note that the majority of the highest probability regions for finding recent, intermediate and ancient infallers are found within the 20$\%$ interlopers contour, meaning the interlopers will not significantly affect the positions of these regions. We will later demonstrate that this is the case directly in Section \ref{sec:result-interloper} and Figure \ref{fig:fig interloper}. In the following section, we extend our results beyond the time since infall, to consider something that has not been quantified so far - what location in phase-space can tell us about the amount of tidal mass loss a galaxy has suffered.

\subsubsection{Probability of a specified strength of tidal mass loss in a projected phase-space diagram}

We also repeat a similar process for the third and fourth column of Figure \ref{fig:fig mas psd}, only this time we calculate the mean probability (and one-sigma error) that a galaxy at that location has suffered tidal stripping of a specified fraction of its halo mass (from top to bottom: increasingly large amounts of tidal mass loss). As before, the grey scale in the third column gives the probability that a galaxy at that location has lost the fraction of halo mass indicated, meanwhile the grey scale in the fourth column gives the one-sigma error on that probability (due to differing line-of-sights, and cluster-to-cluster variations). 

The limits of the halo mass losses are arbitrarily chosen in order to attempt to match location of the peak probabilities in the first column on the same row. However, producing a precise match is challenging as mass loss occurring prior to cluster infall can blur the correlation between time since infall and tidal mass loss (see previous section and Figure \ref{fig:fig relation}). Nevertheless, clearly weak tidal mass loss occurs for galaxies that have spent little (or no) time in the cluster. Meanwhile the strongest mass losses (and highest probabilities) are found in a similar region of phase-space as those haloes which have spent the most time in the cluster (the ancient infallers).

\subsubsection{Projected phase-space diagrams: Separated by region}
\label{sec:result-regional diagram}

In the left panel of Figure \ref{fig:fig region psd}, we divide the projected phase-space diagram, of all the groups/clusters combined, into five regions (labelled A-E). The choice of each region's location and shape is somewhat arbitrary, but is designed to try to maximize the fraction of galaxies belonging to a particular time since infall group in any region (see right panel), while filling the volume occupied by cluster galaxies in projected phase-space. We use the contours in Figure \ref{fig:fig mas psd} to give us good educated guesses on their location and shape. Region C does not match the location of the contours in Figure \ref{fig:fig mas psd}, but is instead designed to span the mixing area between region B and D, so as region B and D are more dominated by the recent infallers and intermediate infallers groups, respectively. By separating the projected phase-space diagram into well-defined regions, and giving a clear break down of the properties of each region, we hope to provide an intuitive, and user-friendly guide to what phase-space diagrams can reveal about time since infall, and tidal mass loss, that can be directly applied to observations of groups and clusters. We note that, as commented in \cite{Ometal13}, in the region partially containing back-splash galaxies (e.g., region D), there are galaxies with a large range of times since infall. It maybe possible to use other indicators, such as gas fraction, to identify true back-splash galaxies \citep[as demonstrated in H.][]{Yoetal16}. The fraction of interlopers (see Section \ref{sec:interlopersintro} for their definition) in each region is represented by cyan colored bars in the right hand panel. With the exception of region A, there are few interlopers in any of the other regions. We note that the information provided in Figure \ref{fig:fig region psd} is provided as supplementary data.

%%%%%%%%%%%%%%%%%%%%%%%%%%%%%%%%%%%%%%%%%%%%%%%%%%%%%%%%%%%%%%%%%%%%
%%%%%%%%%		Figure 7		%%%%%%%%%%%%%%%%%%%%%%%%%%%%%%%%%%%%%%%%%%%%%%
\begin{figure}																			%
\centering 																			%
\includegraphics[width=0.49\textwidth]{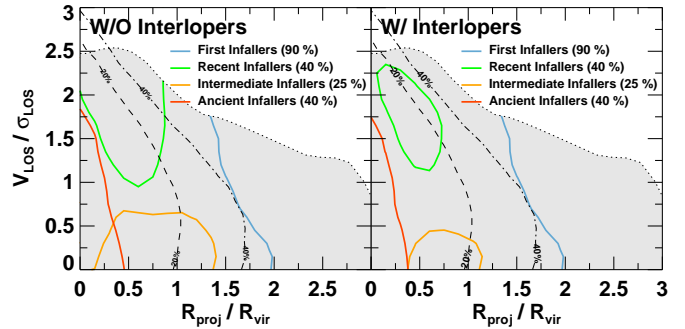}												%
\caption{Distributions of time since infall populations with and without interlopers. Colored lines are isocontours of different populations to trace peak regions with arbitrary values. Again, black dashed and black dot-dashed lines represent contours of interloper fractions. Between two panels, distributions are remaining with small changes.}																	%	
\label{fig:fig interloper}																	%
\end{figure}																			%
%%%%%%%%%		End Figure 	%%%%%%%%%%%%%%%%%%%%%%%%%%%%%%%%%%%%%%%%%%%%%%
%%%%%%%%%%%%%%%%%%%%%%%%%%%%%%%%%%%%%%%%%%%%%%%%%%%%%%%%%%%%%%%%%%%%

%%%%%%%%%%%%%%%%%%%%%%%%%%%%%%%%%%%%%%%%%%%%%%%%%%%%%%%%%%%%%%%%%%%%
%%%%%%%%%		Figure 8		%%%%%%%%%%%%%%%%%%%%%%%%%%%%%%%%%%%%%%%%%%%%%%
\begin{figure*}																			%
\centering 																			%
\includegraphics[width=0.95\textwidth]{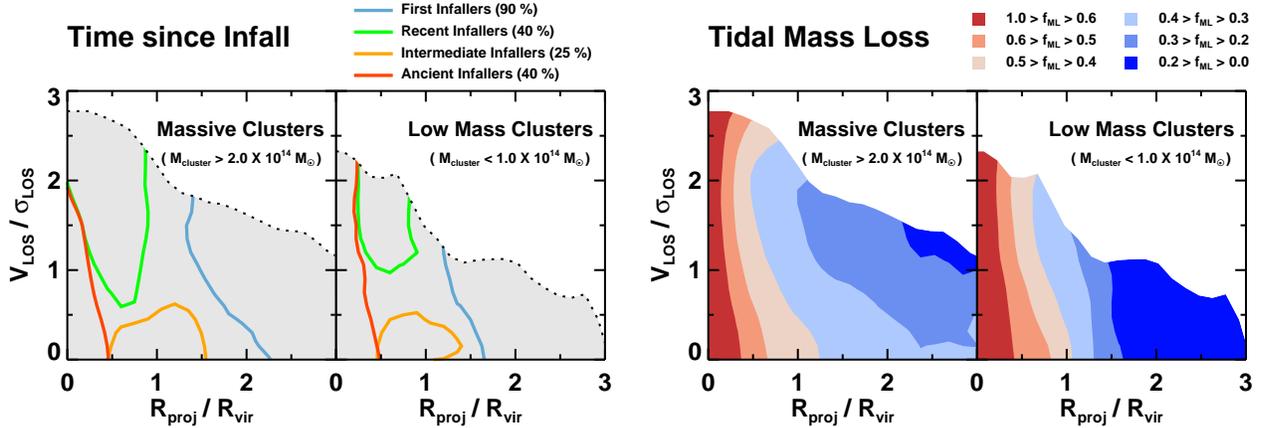}												%
\caption{Dependency of phase-space predictions on group/cluster mass measured at $z$$=$$0$. Left panel: for time since infall, right panel: for mean tidal mass loss. Colored lines are contours to show where a particular time since infall group is found in fractions above the percentage given in the legend. The color of filled regions indicates the bin of mean tidal mass loss (see legend, right panel). Grey shading indicates the total region inhabited by galaxies following cluster rotations. The effect of interlopers are not considered in this figure.}																	%
\label{fig:fig contour cluster}																%
\end{figure*}																			%
%%%%%%%%%		End Figure 	%%%%%%%%%%%%%%%%%%%%%%%%%%%%%%%%%%%%%%%%%%%%%%
%%%%%%%%%%%%%%%%%%%%%%%%%%%%%%%%%%%%%%%%%%%%%%%%%%%%%%%%%%%%%%%%%%%%

\subsubsection{The effects of interlopers in projected phase-space}
\label{sec:result-interloper}

As described in Section \ref{sec:interlopersintro}, our hydrodynamical simulations are only zoomed to 3 virial radii. Therefore, projected phase-space diagrams produced with their data, alone, do not include interlopers from beyond 3 virial radii. Therefore we have supplemented our hydrodynamical simulations with a large scale dark matter only cosmological simulation which we use to quantify the impact of the missing interlopers on our results.

The contours in Figure \ref{fig:fig mas psd}, and the cyan bars in Figure \ref{fig:fig region psd} illustrate that that the number of interlopers is quite low in the regions of phase-space that are of particular interest (e.g., the regions where the highest probability of finding recent, intermediate and ancient infallers are found). To demonstrate this directly, we present Figure \ref{fig:fig interloper}, where the impact of the absence (left panel) or presence (right panel) of interlopers on the probability contours of a particular time since infall population can be directly compared.

The contour of ancient infallers is almost unaffected due to the small quantity of interlopers in that region of phase-space (i.e., the ancient infallers are deep within the 20$\%$ probability contour of the interlopers shown by a black dashed line). Both the recent and intermediate infallers regions are reduced in size, as the presence of additional interlopers within that region has slightly reduced the probabilities of finding each time since infall population. However, in general, the overall location of the regions is quite similar. The contour of first infallers is identical in both plots by construction as both first infallers and interlopers do not have a time since infall yet.

In addition, we note that the way we have considered interlopers assumes that an observer has no knowledge about whether a galaxy is truly a cluster member or not, beyond the projected radius and line-of-sight velocity. As a result, galaxies from as much as $\sim$10 virial radii from the cluster may appear within a projected phase-space diagram. This maybe the case for distant clusters, but for some nearby clusters there may be alternative means to establish cluster membership, or at least distinguish very distant galaxies, such as galaxy morphology \citep{Bietal85}, galaxy color \citep{Yoetal08}, HI content \citep{Chetal07}, surface brightness fluctuations \citep{Meetal07}, the Tully-Fisher relations and/or the Fundamental plane \citep{Gaetal99}. Therefore the effect shown in Figure \ref{fig:fig interloper} may represent the worst case scenario for the effect of interlopers on phase-space diagrams, in the complete absence of any other indicator of a galaxy's location with respect to the cluster.

%%%%%%%%%%%%%%%%%%%%%%%%%%%%%%%%%%%%%%%%%%%%%%%%%%%%%%%%%%%%%%%%%%%%
%%%%%%%%%		Figure 9		%%%%%%%%%%%%%%%%%%%%%%%%%%%%%%%%%%%%%%%%%%%%%%
\begin{figure*}																			%
\centering 																			%
\includegraphics[width=0.95\textwidth]{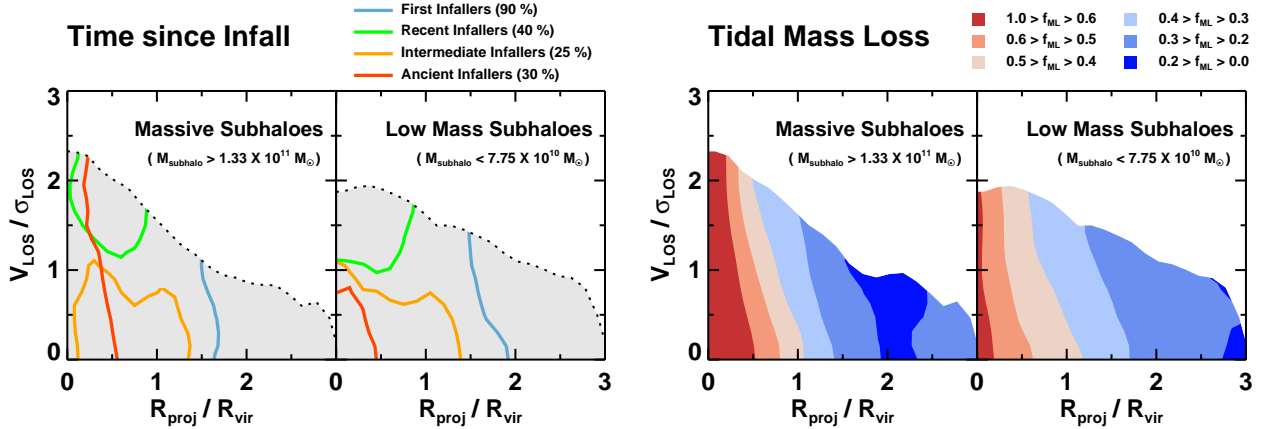}												%
\caption{Same as Figure \ref{fig:fig contour cluster} but now testing the dependency of projected phase-space diagrams on galaxy mass measured at peak mass. We decrease the contour value of the ancient infallers to 30$\%$ compared to previous plots to better trace the ancient infallers in the low mass subhalo sample. The effect of interlopers are not considered in this figure.}																			%
\label{fig:fig contour subhalo}																%
\end{figure*}																			%
%%%%%%%%%		End Figure 	%%%%%%%%%%%%%%%%%%%%%%%%%%%%%%%%%%%%%%%%%%%%%%
%%%%%%%%%%%%%%%%%%%%%%%%%%%%%%%%%%%%%%%%%%%%%%%%%%%%%%%%%%%%%%%%%%%%

%%%%%%%%%%%%%%%%%%%%%%%%%%%%%%%%%%%%%%%	Result-Phase space diagram dependency	%%%%%%%%%%%%%%%%%%%%%%%%%%%%%%%%%%%%%%%%%%%%%%%%

\subsection{Dependencies on cluster mass and galaxy mass}
\label{sec:result-psd dependency}

\subsubsection{Cluster Mass Dependency}
\label{sec:result-psd dependency-cluster}

We now test if the distribution of galaxies in projected phase-space depend on the mass of the group/cluster. We build two subsamples, based on their virial mass at $z$$=$$0$ (equivalent to their peak halo mass); \textit{massive clusters} ($\rm{M}_{\rm cluster}$$>$$2.0\times10^{14}\,\rm{M}_{\odot}$) and \textit{low mass clusters} ($\rm{M}_{\rm cluster}$$<$$1.0\times10^{14}\,\rm{M}_{\odot}$) and re-build their projected phase-space diagrams (see the Figure \ref{fig:fig contour cluster}). ``Massive'' clusters show the clusters C1 through C5 and ``low-mass'' clusters show C11 through C15 in Table \ref{tab:all_clusters}. The left two panels relate to time since infall, and use contours to compare the distributions. The right two panels use colored regions (see associated legend) to compare the mean tidal mass loss. The light grey area just delimits where galaxies are found after combining them together for these subsamples. We note that the effect of interlopers is not considered in this plot, as the focus is on the comparison between the two subsamples.

We find that the ancient and intermediate infallers are distributed quite similarly, independent of cluster mass. {\it{This means that the projected phase-space diagrams in this study can be safely used to predict the location of galaxies that fell into the cluster more than 3.6 Gyr ago, independent of whether the observed galaxies are in a group or a cluster.}} However, the shape of the first infallers contour is more curved in the massive clusters. This is likely because of the effects of the back-splash populations, whose greater presence in the massive cluster subsample forces the first infallers contour to large radius at low velocities. One possible explanation could be that the dynamical friction of satellites in the more massive clusters is less effective due to the greater mass difference between them, \citep[as discussed in][]{Ometal13}. However, when we split our sample in two subsamples by mass ratio, we could not find any significant difference between the two subsamples. Another potential difference could be the growth rate/merger history of the groups/clusters. However, we will leave a detailed analysis of the back-splash population to a dedicated follow-up paper. The recent infallers show a slightly different shape in the two subsamples -- the distribution is wider in the more massive clusters. We confirm that the fraction of recent infallers compared to the total population is very similar in both subsamples, as is the distribution of their times since infall. However, we find that if we had chosen to use the 30$\%$ contour (instead of the 40$\%$ contour), that the recent infallers were distributed very similarly. Therefore it is possible that the differences in shape could simply arise due to low number statistics. Nevertheless, the location of the recent infallers is very similar in both low and high mass clusters.

In the right panels, we find that the distribution of tidal mass loss is fairly similar within one virial radius, independent of the group/cluster mass. Significant differences only appear beyond one virial radius and at small velocities, and are once again likely to be related to the more prevalent presence of the back-splash population in the massive cluster subsample.

\subsubsection{Dependency on Galaxy Halo Mass}
\label{sec:result-psd dependency-halo}

Now we split our total sample of galaxies into a low mass sample ($<$$7.75\times10^{10}$M$_{\odot}$), and a high mass sample ($>$$1.33\times10^{11}$M$_{\odot}$), based on their peak halo mass. Each subsample contains about 40\%\ of total galaxies. We then build projected phase-space diagrams for the subsamples (see Figure \ref{fig:fig contour subhalo}), where the layout and design of the figure panels is identical to that of Figure \ref{fig:fig contour cluster}. The effect of interlopers is not considered in this plot, as the focus is on the comparison between the two subsamples.

We find that the distribution of the time since infall groups, and the tidal mass loss groups to be generally similar, independent of galaxy mass. The contours of first infallers, recent infallers and intermediate infallers are very similar. However, there appears to be a deficit of ancient infallers at high velocities in the low mass subsample. The fraction of ancient infallers compared to the total sample, and their distribution of time since infall, is very similar in the two subsamples. We find that if we had instead chosen to plot the 30$\%$ contour (instead of the 40$\%$ contour), the deficit would have disappeared. Therefore it is possible that it simply arises due to low number statistics, for example, if the recent infallers are by chance found at smaller radii in the low mass subsample. In any case, the locations of the contours are generally very similar in both subsamples, implying they are nearly	 independent of galaxy halo mass. {\it{This suggests that the projected phase-space diagrams we have produced in this study can be applied generally to galaxies observed in clusters without needing to consider differing galaxy masses.}} We reemphasize an earlier point that, in these simulations, we have excellent mass resolution. Our lowest halo mass halo is $3\times10^{10}$M$_{\odot}$ which, assuming halo abundance matching, corresponds to a galaxy stellar mass of about $10^{7}$M$_{\odot}$.
 
%%%%%%%%%%%%%%%%%%%%%%%%%%%%%%%%%%%%%%%%%%%%%%%%%%%%%%%%%%%%%%%%%%%%%%%%%%%%%%%%%%%%%%%%%%%%%%%%%%%%%%%%%%%%%%%%%%%
%%%%%%%%%%%%%%%%%%%%%%%%%%%%%%%%%%%%%%%%	Conclusion & Summary	%%%%%%%%%%%%%%%%%%%%%%%%%%%%%%%%%%%%%%%%%%%%%%%%%%%%%%%%%%
%%%%%%%%%%%%%%%%%%%%%%%%%%%%%%%%%%%%%%%%%%%%%%%%%%%%%%%%%%%%%%%%%%%%%%%%%%%%%%%%%%%%%%%%%%%%%%%%%%%%%%%%%%%%%%%%%%%

\section[]{Conclusion and Summary}
\label{sec:con and sum}

Using cosmological hydrodynamic simulations of groups and clusters with excellent mass resolution, and a full baryonic physics treatment including supernova and AGN feedback, we test what insights we can gain about environmental effects based on the location of galaxies in phase-space diagrams (plots of velocity versus radius, with respect to the group/cluster). We confirm previous authors' results, that is, if the full three-dimensional velocity and three-dimensional clustocentric radius is known, galaxies that fall in at different times tend to be found in particular regions of phase-space (see Figure \ref{fig:fig 3d psd}). In addition, we find that there is a direct correlation between time since infall and tidal mass loss. This means that location in phase-space also provides information on the amount of tidal mass loss a galaxy has suffered, although there is some spread from the correlation due to pre-processing effects. However, typically observations of samples of galaxies in clusters have line-of-sight velocities, and projected radii measurements. Therefore we instead build {\it{user-friendly projected phase-space diagrams}} with the simulated cluster galaxies, so as our simulated results can be directly compared with the observations, accounting for differing line-of-sights, and cluster-to-cluster variations. We also provide the data in our main plots as supplementary data to try to encourage the use of phase-space diagrams as a tool for understanding environmental effects. Our key conclusions are as follows:

\begin{itemize}
\item Galaxies with different times since infall into the cluster preferentially occupy different locations in projected phase-space diagrams (see the first column of Figure \ref{fig:fig mas psd}). We quantify the probability that a galaxy in a particular location belongs to a particular infall group. We also quantify the one-sigma error on this probability due to cluster-to-cluster variations, and differing lines-of-sight (see the second column of Figure \ref{fig:fig mas psd}).
\item The amount of tidal mass loss of dark matter that arises from the cluster potential increases nearly linearly, on average, with the time a galaxy has spent in the cluster (see Figure \ref{fig:fig relation}). We find that tidal mass loss after the first pericenter passage is the origin for this trend (see Figure \ref{fig:fig peritest}). Because of this trend, the fact that location in projected phase-space diagrams provides information on time since infall, means that they also provide information on amount of tidal mass loss (note the similarity between the first and third column of Figure \ref{fig:fig mas psd}).
\item We provide plots that can be directly and easily compared with observed samples of cluster galaxies. For example, we provide a breakdown of the different regions in projected phase-space, and the typical time since infall and tidal mass loss in each region (see Figure \ref{fig:fig region psd}).
\item We find that the predictions for time since infall, and tidal mass loss of dark matter from projected phase-space diagrams do not vary strongly with cluster mass, and/or galaxy mass. This means that the plots we provide can be applied to groups and clusters alike, and independent of the mass of the galaxies (see Figure \ref{fig:fig contour cluster} and Figure \ref{fig:fig contour subhalo}).\footnote{We caution that, while this is true for tidal stripping, ram pressure stripping maybe much more sensitive to galaxy and cluster mass \citep[for example see][]{Jaetal15}.}
\end{itemize}

This study demonstrates that projected phase-space diagrams can provide an essential tool for improving our understanding of environmental effects acting in groups and clusters of galaxies. However, we caution against the use of phase-space diagrams on a galaxy-by-galaxy basis. Clearly any individual galaxy's line-of-sight velocity may be only a fraction of their true orbital velocity. Similarly, their projected distance is a lower limit on their true, three-dimensional clustocentric radius. Instead, the value of phase-space diagrams is greatly increased when dealing with statistically valid, large samples of galaxies.

We also note that to build a phase-space diagram, we have to choose a single location for the cluster center, and all velocities and distances are then measured with respect to that center. Clearly in the case of highly irregular clusters, for example, where two subclusters of equal mass are in the process of merging, then the approximation of a single cluster center could cause substantial scatter of the phase-space distribution. In addition, it is challenging to define the center of a group with low mass and few satellites. Furthermore, we have already shown that galaxies that lost mass in environments outside the cluster (e.g., in galaxy groups) can weaken the link between time since infall into the cluster, and tidal mass loss (see Figure \ref{fig:fig relation}). Therefore in the future we will investigate ways to reduce the one-sigma errors in our predicted times since infall, and tidal mass loss fractions, by reducing the impact of substructure, using approaches that can be directly applied to observed clusters.

%%%%%%%%%%%%%%%%%%%%%%%%%%%%%%%%%%%%%%%%%%%%%%%%%%%%%%%%%%%%%%%%%%%%%%%%%%%%%%%%%%%%%%%%%%%%%%%%%%%%%%%%%%%%%%%%%%%-
%%%%%%%%%%%%%%%%%%%%%%%%%%%%%%%%%%%%%%%%	Acknowledgement	%%%%%%%%%%%%%%%%%%%%%%%%%%%%%%%%%%%%%%%%%%%%%%%%%%%%%%%%%%%%%
%%%%%%%%%%%%%%%%%%%%%%%%%%%%%%%%%%%%%%%%%%%%%%%%%%%%%%%%%%%%%%%%%%%%%%%%%%%%%%%%%%%%%%%%%%%%%%%%%%%%%%%%%%%%%%%%%%%
\acknowledgements
R.S. and S.K.Y. acknowledges support from the Korean National Research Foundation (NRF-2017R1A2A1A05001116), and from Brain Korea 21 Plus Program(21A20131500002). This study was performed under the umbrella of the joint collaboration between Yonsei University Observatory and the Korean Astronomy and Space Science Institute. As head of the group, S.K.Y acted as a corresponding author. YJ was co-funded by the Marie Curie Actions of the European Commission (FP7-COFUND). The supercomputing time for the numerical simulations was kindly provided by KISTI (KSC-2014-G2-003). This study was within the activity of the joint collaboration between YUO and KASI.

%%%%%%%%%%%%%%%%%%%%%%%%%%%%%%%%%%%%%%%%%%%%%%%%%%%%%%%%%%%%%%%%%%%%%%%%%%%%%%%%%%%%%%%%%%%%%%%%%%%%%%%%%%%%%%%%%%%
%%%%%%%%%%%%%%%%%%%%%%%%%%%%%%%%%%%%%%%%	Reference		%%%%%%%%%%%%%%%%%%%%%%%%%%%%%%%%%%%%%%%%%%%%%%%%%%%%%%%%%%%%%%%%
%%%%%%%%%%%%%%%%%%%%%%%%%%%%%%%%%%%%%%%%%%%%%%%%%%%%%%%%%%%%%%%%%%%%%%%%%%%%%%%%%%%%%%%%%%%%%%%%%%%%%%%%%%%%%%%%%%%

%%%%%%%%%%%%%%%%%%%%%%%%%%%%%%%%%%%%%%%%%%%%%%%%%%%%%%%%%%%%%%%%%%%%%%%%%%%%%%%%%%%%%%%%%%%%%%%%%%%%%%%%%%%%%%%%%%%
%%%%%%%%%%%%%%%%%%%%%%%%%%%%%%%%%%%%%%%%	End		%%%%%%%%%%%%%%%%%%%%%%%%%%%%%%%%%%%%%%%%%%%%%%%%%%%%%%%%%%%%%%%%%%%
%%%%%%%%%%%%%%%%%%%%%%%%%%%%%%%%%%%%%%%%%%%%%%%%%%%%%%%%%%%%%%%%%%%%%%%%%%%%%%%%%%%%%%%%%%%%%%%%%%%%%%%%%%%%%%%%%%%
\clearpage

\begin{thebibliography}{0}%
\makeatletter
\providecommand \@ifxundefined [1]{%
 \@ifx{#1\undefined}
}%
\providecommand \@ifnum [1]{%
 \ifnum #1\expandafter \@firstoftwo
 \else \expandafter \@secondoftwo
 \fi
}%
\providecommand \@ifx [1]{%
 \ifx #1\expandafter \@firstoftwo
 \else \expandafter \@secondoftwo
 \fi
}%
\providecommand \natexlab [1]{#1}%
\providecommand \enquote  [1]{``#1''}%
\providecommand \bibnamefont  [1]{#1}%
\providecommand \bibfnamefont [1]{#1}%
\providecommand \citenamefont [1]{#1}%
\providecommand \href@noop [0]{\@secondoftwo}%
\providecommand \href [0]{\begingroup \@sanitize@url \@href}%
\providecommand \@href[1]{\@@startlink{#1}\@@href}%
\providecommand \@@href[1]{\endgroup#1\@@endlink}%
\providecommand \@sanitize@url [0]{\catcode `\\12\catcode `\$12\catcode
  `\&12\catcode `\#12\catcode `\^12\catcode `\_12\catcode `\%12\relax}%
\providecommand \@@startlink[1]{}%
\providecommand \@@endlink[0]{}%
\providecommand \url  [0]{\begingroup\@sanitize@url \@url }%
\providecommand \@url [1]{\endgroup\@href {#1}{\urlprefix }}%
\providecommand \urlprefix  [0]{URL }%
\providecommand \Eprint [0]{\href }%
\providecommand \doibase [0]{http://dx.doi.org/}%
\providecommand \selectlanguage [0]{\@gobble}%
\providecommand \bibinfo  [0]{\@secondoftwo}%
\providecommand \bibfield  [0]{\@secondoftwo}%
\providecommand \translation [1]{[#1]}%
\providecommand \BibitemOpen [0]{}%
\providecommand \bibitemStop [0]{}%
\providecommand \bibitemNoStop [0]{.\EOS\space}%
\providecommand \EOS [0]{\spacefactor3000\relax}%
\providecommand \BibitemShut  [1]{\csname bibitem#1\endcsname}%
\let\auto@bib@innerbib\@empty
%</preamble>
\end{thebibliography}%


\begin{thebibliography}{}
\bibitem[Abadi et al.(1999)]{Abetal99} Abadi, M. G., Moore, B., Bower, R. G. 1999, MNRAS, 308, 947
\bibitem[Aubert et al.(2004)]{Auetal04} Aubert, D., Pichon, C., \&\ Colombi, S. 2004, MNRAS, 352, 376
\bibitem[Bekki(2009)]{Be09} Bekki, K. 2009, MNRAS, 399, 2221
\bibitem[Bertschinger(1985)]{Be85} Bertschinger, E. 1985, ApJS, 58, 39
\bibitem[Binggeli et al.(1985)]{Bietal85} Binggeli, B., Sandage, A., Tammann, G. A. 1985, AJ, 90, 1681
\bibitem[Boselli et al.(2008)]{Boetal08} Boselli, A., Boissier, S., Cortese, L., Gavazzi, G. 2008, ApJ, 674, 742
\bibitem[Byrd \&\ Valtonen(2001)]{BV01} Byrd, G., Valtonen, M. 2001, AJ, 121, 2943
\bibitem[Choi \&\ Yi.(2017)]{Chetal17} Choi, H., Yi, S. 2017, ApJ, 837, 68
\bibitem[Chung et al.(2007)]{Chetal07} Chung, A., van Gorkom, J. H., Kenney, J. D. P., Vollmer, B. 2007, ApJ, 659, 115
\bibitem[De Lucia et al.(2012)]{Deetal12} De Lucia, G., Weinmann, S., Poggianti, B. M., Arag\'{o}n-Salamanca, A., Zaritsky, D. 2012, MNRAS, 423, 1277
\bibitem[Dressler(1980)]{Dr80} Dressler, A. 1980, ApJ, 629, 143
\bibitem[Dubois et al.(2016)]{Duetal16} Dubois, Y., Peirani, S., Pichon, C., Devriendt, J. Gavazzi, R., Welker, C., Volonteri, M. 2016, MNRAS, 463, 3948
\bibitem[Dubois et al.(2014)]{Duetal14} Dubois, Y., Pichon, C., Welker, C., Le Borgne, D., Devriendt, J., et al. 2014, MNRAS, 444, 1453
\bibitem[Dubois et al.(2012)]{Duetal12} Dubois, Y., Devriendt, J., Slyz, A., \&\ Teyssier, R. 2012, MNRAS, 420, 2662
\bibitem[Dubois \&\ Teyssier(2008)]{DT08} Dubois, Y., \&\ Teyssier, R. 2008, A\&A, 477, 79
\bibitem[Font et al.(2008)]{Foetal08} Font, A. S., Bower, R. G., McCarthy, I. G., et al. 2008, MNRAS, 389, 1619
\bibitem[Gao et al.(2004)]{Gaetal04} Gao, L., White, S. D. M., Jenkins, A., Stoehr, F., Springel, V. 2004, MNRAS, 355, 819
\bibitem[Gavazzi et al.(1999)]{Gaetal99} Gavazzi, G., Boselli, A., Scodeggio, M., Pierini, D., Belsole, E. 1999, MNRAS, 304, 595
\bibitem[Gill et al.(2005)]{Gietal05} Gill, S. P. D., Knebe, A., Gibson, B. K. 2005, MNRAS, 356, 1327
\bibitem[Gill et al.(2004b)]{Gietal04b} Gill, S. P. D., Knebe, A., Gibson, B. K., Dopita, M. A. 2004b, MNRAS, 351, 410
\bibitem[Gill et al.(2004a)]{Gietal04a} Gill, S. P. D., Knebe, A., Gibson, B. K. 2004, MNRAS, 351, 399
\bibitem[Gnedin et al.(2003)]{Gnetal03} Gnedin, O. Y. 2003, ApJ, 589, 752
\bibitem[Gonz\'{a}lez-Garc\'{i}a et al.(2005)]{Goetal05} Gonz\'{a}lez-Garc\'{i}a, A. C., Aguerri, J. A. L., Balcells, M. 2005, A\&A, 444, 803
\bibitem[Gunn \&\ Gott(1972)]{GG72} Gunn, J. E., Gott, J. R. I. 1972, ApJ, 176, 1
\bibitem[Guo et al.(2010)]{Guetal10} Guo, Q., White, S., Li, C., Boylan-Kolchin, M. 2010, MNRAS, 404, 1111
\bibitem[Haardt \&\ Madau(1996)]{HM96} Haardt, F., \&\ Madau, P. 1996, ApJ, 461, 20
\bibitem[Haines et al.(2015)]{Haetal15} Haines, T., McIntosh, D. H., S\'{a}nchez, S. F., Tremonti, C., Rudnick, G. 2015, MNRAS, 451, 433
\bibitem[Hern\'{a}ndez-Fern\'{a}ndez et al.(2014)]{HFetal14} Hern\'{a}ndez-Fern\'{a}ndez, J. D., Haines, C. P., Diaferio, A., Iglesias-P\'{a}ramo, J., Mendes de Oliveira, C., Vilchez, J. M. 2014, MNRAS, 438, 2186
\bibitem[Hou et al.(2014)]{Hoetal14} Hou, A., Parker, L. C., Harris, W. E. MNRAS, 442, 406
\bibitem[J\'{a}chym et al.(2007)]{Jaetal07} J\'{a}chym, P., Palou\v{s}, J., K\"{o}ppen, J., Combes, F. A\&A, 472, 5
\bibitem[Jaff\'{e} et al.(2016)]{Jaetal16} Jaff\'{e}, Y. L., Verheijen, M. A. W., Haines, C. P., Yoon, H., Cybulski, R., Montero-Casta\~{n}o, M., Smith, R., Chung, A., Deshev, B. Z., Fern\'{a}ndez, X., van Gorkom, J., Poggianti, B. M., Yun, M. S., Finoguenov, A., Smith, G. P., Okabe, N. 2016, MNRAS, 461, 1202
\bibitem[Jaff\'{e} et al.(2016)]{Jaetal16} Jaff\'{e}, Y. L., Verheijen, M. A. W., Haines, C. P., et al. 2016, MNRAS, 461, 1202
\bibitem[Jaff\'{e} et al.(2015)]{Jaetal15} Jaff\'{e}, Y. L., Smith, R., Candlish, G. N., Poggianti, B. M., Sheen, Y., Verheijen, M. A. 2015, MNRAS, 448, 1715
\bibitem[Klypin et al.(1999)]{Kletal99} Klypin, A., Gottl\"{o}ber, S., Kravtsov, A. V., Khokhlov, A. M. 1999, ApJ, 516, 530
\bibitem[Komatsu et al.(2011)]{Koetal11} Komatsu, E., Smith, K. M., Dunkley, J., et al 2011, The Astrophysical Journal Supplement Series, 192, 18
\bibitem[Larson et al.(1980)]{Laetal80} Larson, R. B., Tinsley, B. M., Caldwell, C. N. 1980, ApJ, 237, 692
\bibitem[Limousin et al.(2009)]{Lietal09} Limousin, M., Sommer-Larsen, J., Natarajan, P., Milvang-Jensen, B. 2009, ApJ, 696, 1771
\bibitem[Lokas et al.(2016)]{Loetal16} Lokas, E. L., Ebrov\'{a}, I., del Pino, A., Sybilska, A., Anthanassoula, E., Semczuk, M., Gajda, G., Fouquet, S. 2016, ApJ, 826, 227
\bibitem[Navarro, Frenk \&\ White(1996)]{NFW96} Navarro, J. F., Frenk, C. S., White, S. D. M. 1996, ApJ, 462, 563
\bibitem[Mahajan et al.(2011)]{Maetal11} Mahajan, S., Mamon, G. A., Raychaudhury, S. 2011, MNRAS, 418, 2816
\bibitem[Mastropietro et al.(2005)]{Maetal05} Mastropietro, C., Moore, B., Mayer, L., Debattista, V. P., Piffaretti, R., Stadel, J. 2005, MNRAS, 364, 607
\bibitem[McCarthy et al.(2008)]{Mcetal08} McCarthy, I. G., Frenk, C. S., Font, A. S., et al. 2008, MNRAS, 383, 593
\bibitem[McGee et al.(2009)]{Mcetal09} McGee, S. L., Balogh, M. L., Bower, R. G., Font, A. S., McCarthy, I. G. 2009, MNRAS, 400, 937
\bibitem[Mei et al.(2007)]{Meetal07} Mei, S., Blakeslee, J. P., C\^{o}t\'{e}, P., et al. 2007, ApJ, 655, 144
\bibitem[Mihos(2004)]{Mi2004} Mihos, J. C. 2004, in Cluster of Galaxies: Probes of Cosmological Structure and Galaxy Evolution, ed. J. S. Mulchaey, A. Dressler, \&\ A. Oemler, 277-+
\bibitem[Moore et al.(1999)]{Moetal99} Moore, B., Lake, G., Quinn, T., Stadel, J. 1999, MNRAS, 304, 465
\bibitem[Moore et al.(1998)]{Moetal98} Moore, B., Lake, G., Katz, N. 1998, ApJ, 495, 139
\bibitem[Muzzin et al.(2014)]{Muetal14} Muzzin, A., van der Burg, R. F. J., McGee, S. L., et al. 2014, ApJ, 796, 65
\bibitem[Oman \&\ Hudson(2016)]{Ometal16} Oman, K, A., Hudson, M. J. 2016, MNRAS, 463, 3083
\bibitem[Oman et al.(2013)]{Ometal13} Oman, K. A., Hudson, M. J., Behroozi, P. S. 2013, MNRAS, 431, 2307
\bibitem[Peirani et al.(2016)]{Peetal16} Peirani, S., Dubois, Y., Volonteri, M., Devriendt, J., Bundy, K., Silk, J., Pichon, C., Kaviraj, S., Gavazzi, R., Habouzit, M. 2016, MNRAS, https://arxiv.org/abs/1611.09922
\bibitem[Prenet et al.(2008)]{Pretal08} Prunet, S., Pichon, C., Aubert, D., et al. 2008, ApJS, 178, 179
\bibitem[Semczuk et al.(2017)]{Seetal17} Semczuk, M., Lokas, E. L., del Pino, A. 2017, ApJ, 834, 7
\bibitem[Sheen et al.(2012)]{Shetal12} Sheen, Y., Yi, S. K., Ree, C. H., Lee, J. 2012, ApJS, 202, 8
\bibitem[Smith et al.(2016)]{Smetal16} Smith, R., Choi, H., Lee, J., Rhee, J., Sanchez-Janssen, R., Yi, S. K. 2016, ApJ, 833, 109
\bibitem[Smith et al.(2015)]{Smetal15} Smith, R., S\'{a}nchez-Janssen, R., Beasley, M. A., Candlish, G. N., Gibson, B. K., Puzia, T. H., Janz, J., Knebe, A., Aguerri, J. A. L., Lisker, T., Hensler, G., Fellhauer, M., Ferrarese, L., Yi, S. K. 2015, MNRAS, 454, 2502
\bibitem[Smith et al.(2013)]{Smetal13} Smith, R., S\'{a}nchez-Janssen, R., Fellhauer, M., Puzia, T. H., Aguerri, J. A. L., Farias, J. P. 2013, MNRAS, 429, 1066
\bibitem[Smith et al.(2010)]{Smetal10} Smith, R., Davies, J. I., Nelson, A. H. 2010, MNRAS, 405, 1723
\bibitem[Taylor \&\ Babul.(2004)]{TB04} Taylor, J. E., Babul, A. 2004, MNRAS, 348, 811
\bibitem[Teyssier(2002)]{Te02} Teyssier, R. 2002, A\&A, 385, 337
\bibitem[Tweed et al.(2009)]{Twetal09} Tweed D., Devriendt J., Blaizot J., Colombi S., Slyz A.
2009, A\&A, 506, 647
\bibitem[Vijayaraghavan et al.(2013)]{Vietal13} Vijayaraghavan, R., Ricker, P. M. 2013, MNRAS, 435, 2713
\bibitem[Villalobos et al.(2014)]{Vietal14} Villalobos, \'{A}., De lucia, G., Murante, G. 2014, MNRAS, 444, 313
\bibitem[Villalobos et al.(2012)]{Vietal12} Villalobos, \'{A}., De Lucia, G., Borgani, S., Murante, G. 2012, MNRAS, 424, 2401
\bibitem[Vollmer et al.(2001)]{Voetal01} Vollmer, B., Cayatte, V., Balkowski, C., Duschl, W. J. 2001, ApJ, 561, 708
\bibitem[Warnick et al.(2008)]{Waetal08} Warnick, K., Knebe, A., Power, C. 2008, MNRAS, 385, 1859
\bibitem[Wetzel et al.(2013)]{Weetal13} Wetzel, A. R., Tinker, J. L., Conroy, C., van den Bosch, F. C. 2013, MNRAS, 432, 336
\bibitem[Wetzel(2011)]{We2011} Wetzel, A. R. 2011, \mnras, 412, 49
\bibitem[Yi et al.(2013)]{Yietal13} Yi, S. K., Lee, J., Jung, I., Ji, I., Sheen, Y. 2013, A\&A, 554, 122
\bibitem[Yoon et al.(2017, submitted)]{Yoetal16} Yoon, H., Chung, A., Smith, R., Jaff\'{e}, Y. L. submitted
\bibitem[Yoon et al.(2008)]{Yoetal08} Yoon, J. H., Schawinski, K., Sheen, Y., Ree, C. H., Yi, S. K. 2008, ApJS, 176, 414

\end{thebibliography}
\end{document}